\documentclass[aip,jcp,reprint]{revtex4-1}
\usepackage{graphicx}
\usepackage[utf8]{inputenc}
\usepackage{amsmath}
\usepackage{verbatim}
\usepackage{amssymb}
\usepackage{listings}
\usepackage{color}
\usepackage{hyperref}
\usepackage{amsmath}
\usepackage{siunitx}
\usepackage{mathtools}

\usepackage{tikz-cd}

\newcommand\subfig[2]{{Fig.~\ref{#1}{#2}}}
\newcommand\subcap[1]{{(#1)}}

\newcommand{\lifted}[2]{\ensuremath{(#1,#2)}}
\newcommand{\lifting}[3]{\ensuremath{\lifted{#1}{#2} \to \lifted{#1}{#3}}}
\newcommand{\pMet}{p^{\text{Met} }}
\newcommand{\pFact}{p^{\text{Fact} }}
\newcommand{\ZZ}{\mathbb{Z}}

\newcommand{\Ags}{\text{\AA}}
\newcommand{\partpart}[2]{\frac{\partial #1}{\partial #2}}
\newcommand{\partpartshort}[2]{\partial_{#2} #1}
\newcommand{\SET}[1]{\{#1\}}

\newcommand{\fig}[1]{Fig.~\ref{#1}}
\newcommand{\quot}[1]{``#1''}
\newcommand{\tab}[1]{Table~\ref{#1}} 
 
\newcommand{\sect}[1]{Section~\ref{#1}} 
\newcommand{\secttwo}[2]{Sections~\ref{#1} and~\ref{#2}}

\newcommand{\CCAL}{\mathcal{C}}  
\newcommand{\FCAL}{\mathcal{F}}  
\newcommand{\MCAL}{\mathcal{M}}  
\newcommand{\OCAL}{\mathcal{O}}  
\newcommand{\PCAL}{\mathcal{P}}  
\newcommand{\TCAL}{\mathcal{T}}  
\newcommand{\expa}[1]{\mathrm{e}^{#1}}   
\newcommand{\expb}[1]{\exp \glb #1 \grb} 
\newcommand{\expc}[1]{\exp \glc #1 \grc} 
\newcommand{\ranb}[2][]{\ran_{#1} \! \glb #2 \grb}  
\newcommand{\rand}[2][]{\ran_{#1} \! \gld #2 \grd}  

\newcommand{\sinb}[2][]{\sin^{#1} \glb #2 \grb}  
\newcommand{\cosb}[2][]{\cos^{#1} \glb #2 \grb}  

\newcommand{\sinc}[2][]{\sin^{#1} \glc #2 \grc}  

\newcommand{\loga}[2][]{\log^{#1}\! \gla #2 \gra}  
\newcommand{\logb}[2][]{\log^{#1} \glb #2 \grb}  
\newcommand{\logc}[2][]{\log^{#1} \glc #2 \grc}  
\newcommand{\logd}[2][]{\log^{#1} \gld #2 \grd}  

\newcommand{\gla}{\,}  
\newcommand{\gra}{}  
\newcommand{\glb}{\left(}  
\newcommand{\grb}{\right)}  
\newcommand{\glc}{\left[}  
\newcommand{\grc}{\right]}  
\newcommand{\gld}{\left\{}  
\newcommand{\grd}{\right\}}  

\newcommand{\const}{\text{const}}

\newcommand{\TO}{,\ldots,}
\newcommand{\VEC}[1]{\mathbf{#1}}
\newcommand{\dvec}{\VEC{d}}
\newcommand{\evec}{\VEC{e}}

\newcommand{\mvec}{\VEC{m}}
\newcommand{\nvec}{\VEC{n}}

\newcommand{\qvec}{\VEC{q}}
\newcommand{\rvec}{\VEC{r}}

\newcommand{\xvec}{\VEC{x}}

\newcommand{\atilde}{\tilde{a}}

\newcommand{\qtilde}{\tilde{q}}

\newcommand{\Dtilde}{\tilde{D}}

\newcommand{\mean}[1]{\left\langle #1 \right\rangle}
\newcommand{\half}{\frac{1}{2}}

\newcommand\bigO[1]{\ensuremath{\OCAL(#1)}}
\newcommand\bigOb[1]{\ensuremath{\OCAL\glb #1 \grb}}

\newcommand\diff[1]{\mathrm{d}#1}
\DeclareMathOperator{\sincf}{sinc}
\DeclareMathOperator{\ran}{ran}

\newcommand{\maxZeroa}[1]{\gla #1 \gra^+}   
\newcommand{\maxZeroc}[1]{\glc #1 \grc^+}   
\newcommand{\UnitX}{\hat{\evec}_x}
\newcommand{\UnitY}{\hat{\evec}_y}
\newcommand{\UnitZ}{\hat{\evec}_z}
\newcommand{\Cell}{\ensuremath{\CCAL}}
\newcommand{\ActiveCell}{\ensuremath{\CCAL_a}}
\newcommand{\TargetCell}{\ensuremath{\CCAL_t}}
\newcommand{\qCell}{\ensuremath{q^{\text{cell}}}}
\newcommand{\Conf}{\ensuremath{c}}
\newcommand{\OldConf}{\ensuremath{c''}}
\newcommand{\NewConf}{\ensuremath{c'}}
\newcommand{\UpP}{\ensuremath{k^+}}
\newcommand{\DownP}{\ensuremath{k^-}}
\newcommand{\ActiveP}{\ensuremath{a}}
\newcommand{\TargetP}{\ensuremath{t}}
\newcommand{\OneP}{\ensuremath{1}} 
\newcommand{\TwoP}{\ensuremath{2}} 
\newcommand{\ThreeP}{\ensuremath{3}} 
\newcommand{\FourP}{\ensuremath{4}} 
\newcommand{\MassFlow}{\text{mass}} 
\newcommand{\Deltax}{{\eta}}
\newcommand{\KK}{{K}}
\newcommand{\FACTOR}[2]{\glb #1, #2 \grb}
\newcommand{\TYPE}[2][]{\gla \text{#2}\gra_{#1}}
\newcommand{\SETOFM}{\MCAL}
\newcommand{\EVRATE}[2]{q_{#1, #2}}
\newcommand{\FACDERIV}[2]{\qtilde_{#1, #2}}
\newcommand{\TOTEVRATE}{Q}

\newcommand{\COULOMB}{\text{C}} 
\newcommand{\BONDTYPE}{\text{bond}} 
\newcommand{\BENDINGTYPE}{\text{bending}} 
\newcommand{\LJTYPE}{\text{LJ}} 
\newcommand{\COULOMBTYPE}{\text{Coulomb}} 
\newcommand{\PAIRTYPE}{\text{pair}} 
\newcommand{\hoh}{\text{H$_2$O}} 
\newcommand{\BJERR}{l_\text{B}} 
\newcommand{\eq}[1]{eq.~\eqref{#1}}
\newcommand{\eqtwo}[2]{eqs~\eqref{#1} and~\eqref{#2}}
\newcommand{\Eqtwo}[2]{Eqs~\eqref{#1} and~\eqref{#2}}
\newcommand{\Eq}[1]{Eq.~\eqref{#1}}
\newcommand{\dip}{a}
\newcommand{\diptilde}{\atilde}

\begin{document}
\date{\today}
\title{All-atom computations with irreversible Markov chains}

\author{Michael~F.~Faulkner}
\affiliation{H.~H. Wills Physics Laboratory, University of Bristol, Tyndall
Avenue, Bristol BS8 1TL, United Kingdom}
\author{Liang Qin}
\affiliation{Laboratoire de Physique Statistique, D\'{e}partement de physique
de l'ENS, Ecole Normale Sup\'{e}rieure, PSL Research University, Universit\'{e}
Paris Diderot, Sorbonne Paris Cit\'{e}, Sorbonne Universit\'{e}s, UPMC
Univ. Paris 06, CNRS, 75005 Paris, France}
\author{A.~C.~Maggs}
\affiliation{CNRS UMR7083, ESPCI Paris, PSL Research University, 10 rue
Vauquelin, 75005 Paris, France}
\author{Werner Krauth}
\affiliation{Laboratoire de Physique Statistique, D\'{e}partement de physique de l'ENS,
Ecole Normale Sup\'{e}rieure, PSL Research University, Universit\'{e} Paris
Diderot, Sorbonne Paris Cit\'{e}, Sorbonne Universit\'{e}s, UPMC Univ. Paris
06, CNRS, 75005 Paris, France}
\affiliation{Max-Planck-Institut für Physik komplexer Systeme, Nöthnitzer Str. 
38, 01187 Dresden, Germany}

\begin{abstract}
The event-chain Monte Carlo (ECMC) method is an irreversible Markov process
based on the factorized Metropolis filter and the concept of lifted Markov
chains. Here, ECMC is applied to all-atom models of multi-particle interactions
that include the long-ranged Coulomb potential.  We discuss a line-charge model
for the Coulomb potential and demonstrate its equivalence with the standard
Coulomb model with tin-foil boundary conditions.  Efficient factorization
schemes for the potentials used in all-atom water models are presented,
before we discuss the best choice for lifting schemes for factors of more than
three particles.  The factorization and lifting schemes are then applied to
simulations of point-charge and charged-dipole Coulomb gases, as well as to
small systems of liquid water. For a locally charge-neutral system in three
dimensions, the algorithmic complexity is \bigO{N \log N} in the number $N$
of particles. In ECMC, a Particle--Particle method, it is achieved without the
interpolating mesh required for the efficient implementation of other modern
Coulomb algorithms.  An event-driven, cell-veto-based implementation samples
the equilibrium Boltzmann distribution using neither time-step approximations
nor spatial cutoffs on the range of the interaction potentials. We discuss
prospects and challenges for ECMC in soft condensed-matter and biological
physics.
\end{abstract}

\maketitle

\section{Introduction}
\label{sec:Introduction}

\subsection{Irreversible Markov processes}
\label{sec:IntroductionIrreversibleMarkov}

Numerical methods are ubiquitous in the natural sciences, with Markov-chain
Monte Carlo~\cite{Metropolis1953} and molecular dynamics~\cite{Alder1957}
playing central roles.  Markov-chain Monte Carlo applies to any computational
science problem that can be formulated as an (perhaps fictitious)
equilibrium-statistical-physics system and whose solution requires sampling
its probability distribution. As in physical and chemical systems, equilibrium
within the computational context usually means that all probability flows
vanish. This requirement is enforced by the detailed-balance condition, an
essential ingredient of most Markov-chain Monte Carlo methods and notably of
the Metropolis algorithm~\cite{SMAC}.  Monte Carlo algorithms usually take much
time to approach equilibrium~\cite{Levin2008}, and, once in equilibrium, to
generate independent samples. This is, in part, due to the fact that detailed
balance leads to time-reversible Markov-chain dynamics, which is diffusive and
therefore slow.

In recent years, a new class of irreversible \quot{event-chain} Monte Carlo
(ECMC) algorithms has been proposed~\cite{Bernard2009,Michel2014JCP}. ECMC
algorithms violate detailed balance but satisfy a weaker global-balance
condition.  Configurations at large times sample the equilibrium distribution,
but the asymptotic steady state comes with non-vanishing probability
flows. In particle systems with periodic boundary conditions, for example,
atoms may  continue to move preferentially in certain directions. In
continuous spin systems, likewise, configurations realize the  equilibrium
distribution even though spins rotate in a preferred way~\cite{Nishikawa2015,
MichelMayerKrauth2015, Lei2018}.  ECMC moves (displacements of particles,
rotations of spins, etc.) are infinitesimal and persistent: An \quot{active}
particle moves directly from one event to the next, that is, it continues to
move until a proposed move is vetoed by a unique \quot{target} particle, which
in turn becomes the active particle. This passing of the active-particle label
is called a lifting~\cite{Diaconis2000,Chen1999} and this concept overcomes the
characteristic rejections of randomly proposed finite moves in the Metropolis
algorithm. ECMC algorithms are powerful~\cite{Bernard2011, Michel2014JCP,
Nishikawa2015, Lei2018}: In a one-dimensional particle system, they were
demonstrated to mix on shorter time scales than Markov chains that satisfy
detailed balance~\cite{KapferKrauth2017}.

In ECMC, the traditional Metropolis acceptance criterion based on the
change in potential is replaced by a consensus rule. This is the essence
of the factorized Metropolis filter, which applies to translation-invariant
systems with pair-wise interactions between particles~\cite{Michel2014JCP}
and, more generally, to models whose interactions can be split into sets
of independent factors~\cite{Harland2017}.  The ECMC algorithm does not
compute the total system potential energy. This makes it very appealing
for long-range-interacting systems, where this computation is costly.
For Coulomb systems, ECMC altogether avoids traditional algorithms for the
electrostatic potential~\cite{deLeeuw1980-1}, the dominant computational
bottleneck for long-range-interacting models.  Rather, the cell-veto
algorithm~\cite{KapferKrauth2016} efficiently establishes consensus on the
acceptance or the rejection of a proposed move, even if all particles interact
with each other.  This is the starting point for the present work.

Generally, computations in statistical physics fall into two categories. They
either aim at thermodynamic averages (energy, specific heat, spatial
correlation  functions, etc.) or at dynamic properties (time correlations,
nucleation barriers, coarsening, etc.). In principle, the computation of
thermodynamic averages is the realm of Markov-chain Monte Carlo, whereas
the analysis of dynamical behavior calls on molecular dynamics, as it solves
Newton's equations of motion.  Specifically, however, the field of large-scale
all-atom computations with long-ranged interactions is today dominated by
molecular dynamics for both categories. The dominance of molecular dynamics
is rooted in two facts: Firstly, traditional Monte Carlo methods usually
update just \bigO{1} particles at a time, and the acceptation/rejection
step then requires the exact computation of the change in potential.
The best currently known algorithm~\cite{heleneMaggs} for the change in
potential after such a local update in a Coulomb system is of complexity
\bigO{\sqrt{N}} so that one Monte Carlo sweep (a sequential update of all $N$
particles) requires \bigO{N^{3/2}} computations.  In molecular dynamics, in
contrast, the discretized Newton's equations update all particle positions
simultaneously, and the necessary computation of the forces on all particles
comes at a cost of \bigO{N \log N}, much less than for a Monte Carlo sweep.
Secondly, Newtonian dynamics conserves momentum and explores phase space more
efficiently than the local Metropolis algorithm. This advantage of molecular
dynamics over Monte Carlo is, for example, brought out by the different
scaling of the velocity auto-correlation functions in the context of long-time
tails~\cite{AlderWainwrightAutocorrelation1970,WittmerPolinska2011}.

The time evolution of molecular dynamics has physical meaning, but from
an algorithmic point of view, it is constrained by the requirement that it
must implement Newton's law. As a result, there is no additional freedom to
accelerate the exploration of phase space.  In contrast, Monte Carlo dynamics
is non-physical and only constrained by the global-balance condition. A
well-chosen Monte Carlo dynamics can considerably speed up the sampling of
the equilibrium distribution. Those equilibrium samples may also serve as
starting configurations for parallel molecular-dynamics calculations that give
access to high-precision dynamical correlation functions.  Furthermore, if
more complex out-of-equilibrium rare-event physical phenomena (such as protein
folding) are of interest, the timescales of long-time features can be accessed
by the inspection of the rare events produced by parallel simulation on
$N_{\text{proc}}$ processors. Similar to the half-life analysis of radioactive
substances composed of large numbers of atoms, a rare event that takes place on
a time scale $\tau$ on a single processor will then take place on a time scale
$\tau/N_{\text{proc}}$ on one of the $N_{\text{proc}}$ processors.

In this work, we develop the framework for the application of ECMC to classical
long-range-interacting all-atom systems. In particular, we demonstrate
efficient ECMC methods that rigorously sample the canonical ensemble, without
even evaluating the total potential. The factorizations that we implement
with the cell-veto algorithm
allow us to move a single particle from one event to the next in a CPU time
that is independent of the number of point charges in a system. For a locally
charge-neutral system, the mean free path (the mean distance between events)
decreases only logarithmically with the number of point charges in the
system. This implies that the computational effort required to move every
particle in a simulation a constant distance scales as only \bigO{N \log N},
with no approximation and without the numerically intensive interpolation mesh
used in many modern electrostatic simulations.

We validate our algorithm through explicit comparisons with a standard
Metropolis algorithm and with molecular-dynamics simulations, each performed 
with Ewald summations.  We focus on two conceptual issues. One is the 
computation of Coulomb pair-event rates, that is, essentially, the derivatives 
of the two-particle Coulomb potential with respect to the position of the
\quot{active} particle. In the simplest version of ECMC, this corresponds to
the probability with which an active particle will stop and induce a lifting
to a target particle.  The other issue concerns the factorization schemes of
the system potential in which we lump together different interactions that
partially compensate each other so that the ECMC mean free path between
events is much increased.  We first apply our ECMC algorithm to a pair of
like Coulomb point charges and then  to systems of charge-neutral dipoles in a
three-dimensional simulation box with periodic boundary conditions.  We finally
demonstrate the perfect agreement of thermodynamic observables between ECMC and
conventional Monte Carlo and molecular dynamics for up to $256$ water molecules
at standard density and temperature. The ECMC algorithm leaves ample room for
improvements. We expect it to be widely applicable to all-atom simulations of
charged systems.

\subsection{All-atom molecular simulations}
\label{sec:IntroductionMolSimulation}

Of great importance in soft-matter research, biological physics and
related fields, the all-atom approach projects the full quantum-mechanical
many-body system onto the reduced classical degrees of freedom of the
atomic positions.  The projection yields the potential energy as a
function of all the particle positions, and the Monte Carlo method can
then, in principle, be applied directly.  Molecular dynamics also starts
from the atomic potential, as the forces in Newton's equations are given
by its spatial derivatives.  Present-day parametrized empirical force-field
models~\cite{leachModelling,schlickModelling} further break up the potentials
and make them amenable to practical computations.  For example, separate
terms in the potential typically describe deviations of chemical bonds from
their equilibrium values, with individual contributions for stretching,
bending and torsion. Likewise, distinct intermolecular potentials capture
longer-ranged features of the interactions; for example, dispersion forces,
hard-core repulsions and long-ranged charge--charge and dipolar interactions.
The all-atom reduction from quantum mechanics to a classical interacting system
is approximate and not uniquely defined.  Various force-field models are used
in a number of code bases~\cite{amberForces,Charmm}, which are also implemented
in other prominent codes~\cite{gromacs,NAMD,ShawAnton}. The parameters in each
force-field model are optimized to reproduce thermodynamic and structural
features over a reduced range of temperatures and pressures. Different
potential functions coexist even for the description of simple molecules such
as water~\cite{compareWater}. We use in this work an all-atom potential for
water that features two-body  bond stretching, three-body bending as well as
long-ranged Coulomb interactions, and a Lennard-Jones potential~\cite{Wu2006}.

Modern codes generally compute the long-ranged Coulomb potential through
variants of the Ewald algorithm applied to a discretized analogue of the
continuous position space.  The Fourier contribution to the potential is
evaluated by first interpolating each point charge to multiple points on
a mesh and then solving the Poisson equation via fast Fourier transform,
which, combined, is of complexity \bigO{N \loga{N}} per computation
of the potential energy. Numerous formulations of this algorithm have
been developed starting with the Particle--Particle--Particle--Mesh
method~\cite{hockney1988ComputerSimulationUsingParticles}. More recent
generations combine the Particle--Mesh philosophy with the Ewald formula, to
create the Particle--Mesh--Ewald method~\cite{meshEwald} together with many
variants~\cite{shawEwald, espresso, multigridEwald} which, together, remain
the workhorse of modern simulation codes. The charge interpolation onto the mesh
generally presents the main computational workload. These methods use intricate
strategies to maintain a high level of accuracy.  Mesh interpolation leads to
very large self-energy artifacts which have to be subtracted with great care in
order not to modify the physical interactions.

Alternative approaches exist for the computation of the Coulomb potential
and the electrostatic forces on particles. The hierarchical multipole-moment
expansion~\cite{greenhard1987}, for example, expands the interactions of a
particle with all the other particles in terms of spherical harmonics, and
therefore avoids Fourier transforms and lattice interpolations. However, the
expansion converges only with high orders of the multipole moments so that
one molecular-dynamics time step, although it is of complexity \bigO{N},
comes with a prohibitive prefactor.  Local algorithms that propagate
electric fields rather than solve the Poisson equation also bypass the
fast Fourier transform~\cite{RossettoPRL,localMC,auxiliary}. This is
an advantage in architectures where the Fourier transform involves
large-scale non-local information transfers.  In these algorithms,
the complexity of a single-particle update is \bigO{1} but the use of a
background lattice to discretize the electrostatic degrees of freedom
again leads to costly interpolations from the continuum charges to
the grid~\cite{RottlerJCP,RottlerPRL}.  In contrast to well-established
methods, ECMC is directly formulated in continuous space, and its successful
implementation only relies on translational invariance on all length scales.
In essence,  ECMC requires no discretization of the simulation box, and
the total Coulomb potential and forces may remain unknown throughout the
simulation.

All-atom molecular-dynamics simulations must take into account a variety
of time scales and lengths. Indeed, the high-precision time integration
of intramolecular spring forces requires a discretization time in the
femtosecond range. The physics associated with the much longer time scales
that one wishes to study include density fluctuations (which relax on
the picosecond time scale), Debye-layer equilibration (nanoseconds), and
conformation changes (milliseconds). At the same time, the precise rendering
of dielectric and screening properties requires high-quality computations,
and the long-ranged nature of the interaction calls for large system sizes in
order to overcome finite-size effects. In order to efficiently manage both
the  stiffness (the presence of many relevant time scales) and long-ranged
potentials, interactions are often broken up, and sophisticated multiple
time-step algorithms are implemented~\cite{timeStepMD, Shaw2010}. Use of
a thermostat~\cite{FrenkelSmitBook2001} is crucial in order to counteract a
drift of the system energy and to connect the potential-energy surface with the
system temperature.  The ECMC algorithm considers the same potentials as its
competitors, but it is fundamentally event-driven so that the exact Boltzmann
distribution is sampled at any given temperature. This renders the thermostat
unnecessary.  In our application, the triggering of events remains well
balanced between intramolecular, short-range intermolecular and long-ranged
intermolecular Coulomb events.

\section{ECMC algorithm}
\label{sec:ECMC}

ECMC~\cite{Bernard2009,Michel2014JCP}  is an irreversible continuous-time
Markov process: Its moves are thus infinitesimal. Analogously, Newton's
differential equations are of course also defined in continuous time.
The molecular-dynamics algorithms that solve Newton's equations must be
time-discretized for all systems except for hard spheres~\cite{Alder1957} or
for stepwise constant potentials~\cite{AlderWainwright1959,BannermanLue2010}.
In contrast, in ECMC, discretization is generally avoided through the
event-driven approach.  In the present section, we discuss the essential issues
of the algorithm's setup and implementation as well as its complexity.

\subsection{Factors, factorized Metropolis filter}
\label{sec:ECMCFactorized}

In ECMC, the interactions in an $N$-particle system are split into a finite or
infinite set of factors $M = \FACTOR{I_M}{T_M} \in \PCAL(\SET{1 \TO N}) \times
\TCAL$, where $\PCAL$ is the power set of the indices (comprising all indices,
pairs of indices, triplets, etc), and $\TCAL$ is a set of interaction types.
We refer to $I_M$ as the index set of the factor and to $T_M$ as its type.
The total potential $U$, which is a function of all particle positions
$\SET{\rvec_1 \TO \rvec_N}$, is written as a sum over factor potentials $U_M$:
\begin{equation}
 U(\SET{\rvec_1 \TO \rvec_N}) = \sum_{M \in \SETOFM}
U_M (\SET{\rvec_i: i \in I_M}), 
\label{equ:PotentialFactorized}
\end{equation}
where $U_M$ only depends on the factor indices $I_M$ and is of type
$T_M$. In \eq{equ:PotentialFactorized}, the set $\SETOFM = \SET{M: U_M
\neq 0} \subset \PCAL(\SET{1 \TO N}) \times \TCAL$ only contains factors
that have a non-zero contribution for some values of the positions. In a
system with only pair interactions, a non-zero factor may be
$\FACTOR{\SET{i,j}}{\TYPE{\PAIRTYPE}}$.  The corresponding factor potential 
would then be $U_{\FACTOR{\SET{i,j}}{\TYPE{\PAIRTYPE}}}(\rvec_{i},\rvec_{j})$, 
and the total potential in \eq{equ:PotentialFactorized} then becomes 
$ U = \sum_{i<j} 
U_{\FACTOR{\SET{i,j}}{\TYPE{\PAIRTYPE}}}(\rvec_{i},\rvec_{j})$, which 
is normally written as $U =  \sum_{i<j} U_\PAIRTYPE(\rvec_i,\rvec_j)$.

In this work, we use more general factorizations.  The Lennard-Jones
factor, that we write as $\FACTOR{\SET{i,j}}{\TYPE{\LJTYPE}}$, has a 
factor potential 
\begin{equation}
U_{\FACTOR{\SET{i,j}}{\TYPE{\LJTYPE}  }}(\rvec_{ij}) = k_{\LJTYPE} \glc \glb 
\frac{\sigma }{ |\rvec_{ij}|} \grb^{12}
- \glb \frac{\sigma }{ |\rvec_{ij}|} \grb ^6 \grc,
\label{equ:LJFACTORINTRO}
\end{equation}
where $\rvec_{ij} = \rvec_j - \rvec_i$ is the shortest separation vector 
from particle $i$ to particle $j$, possibly corrected for periodic boundary 
conditions.
The Lennard-Jones factor
\quot{\LJTYPE} in \eq{equ:LJFACTORINTRO} may be 
replaced by two types,  
namely the type $\TYPE[6]{LJ}$ (describing the $1/| \rvec_{ij}|^6$ part
of the Lennard-Jones interaction) and the type $\TYPE[12]{LJ}$ (describing
its $1/| \rvec_{ij}|^{12}$ part)~\cite{Michel2014JCP}. For two indices $i$
and $j$, this yields two factors, namely $\FACTOR{\SET{i,j}}{\TYPE[6]{LJ}}$
and $\FACTOR{\SET{i,j}}{\TYPE[12]{LJ}}$.  Likewise, the bending energy in a
water molecule with particles $i,j,k$ will correspond to a factor index $I_M =
\SET{i,j,k}$ and to a factor type given by the specific function chosen for
this interaction.  A similar approach was introduced for modeling neighboring
beads in a polymer~\cite{Harland2017}.  In \secttwo{sec:DipoleLiftingSchemes}
{sec:Water}, we consider factors that lump together all of the
Coulomb interactions between the four particles comprising two distinct
dipoles, and even between the six particles of two water molecules,
respectively. The factor corresponding to the latter case is given by
$\FACTOR{\SET{i,j,k,l,m,n}}{\TYPE{Coulomb}}$. (For
simplicity of notation, we do not differentiate in this work the Coulomb types
for two, four and six particles.)  As mentioned, the set of factors can be
infinite~\cite{KapferKrauth2016}, even for finite $N$. As an example, in a
finite periodic system, one can view the three-dimensional Coulomb interaction
between particles $i$ and $j$ as a sum of interactions between $i$ and each
periodic copy of $j$ indexed by an image index $\nvec \in \ZZ^3$. For the case
of the above two-water-molecule Coulomb interaction, we would then have $M =
\FACTOR{\SET{i,j,k,l,m,n}}{\TYPE[\nvec]{Coulomb} }$. The type set $\TCAL$ would
then contain all of the separate-image Coulomb interactions:
\begin{equation}
 \SET{\TYPE[\nvec]{Coulomb}: \nvec \in \ZZ^3} \subseteq \TCAL, 
\end{equation}
where the set of Coulomb types may be a proper or an improper subset of 
$\TCAL$. We will treat such factor types in \sect{sec:Coulomb}.

Given the potential factorization enforced by \eq{equ:PotentialFactorized}, the
Boltzmann weight $\pi(\Conf) = \expc{-\beta U(\Conf)}$ of configuration
$\Conf = \SET{\rvec_1 \TO \rvec_N}$ reduces to a product over factor weights
$\pi_M(\Conf_M) = \expc{-\beta U_M(\Conf_M)}$:
\begin{equation}
\pi(\Conf) = 
\prod_{M} \pi_M(\Conf_M) =
\prod_{M}\expc{-\beta U_{M}(\Conf_M)}, 
\label{equ:BoltzmannFactorized}
\end{equation}
where $\Conf_M$ is the factor configuration, that is, the configuration
$\Conf$ restricted to the indices of factor $M$.  The traditional Metropolis
filter~\cite{Metropolis1953}, which defines the acceptance probability for a
move from configuration $\Conf$ to configuration $\NewConf$ in the Metropolis
algorithm, does not factorize in a similar fashion:
\begin{align}
    \pMet(\Conf \to \NewConf) &= \min\glc 1, \expb{-\beta 
\Delta U} \grc ,
\label{equ:MetropolisProduct}    \\
                                 &= \min\glc 1, \prod_{M} \expb{-\beta 
\Delta U_{M}} 
    \grc ,
\label{equ:MetropolisProduct2}    
\end{align}
where $\Delta U_{M} = U_{M}(\NewConf_M) - U_{M}(\Conf_M)$ is the
factor-potential difference between factor configurations $\Conf_M$ and
$\NewConf_M$.  The recent factorized Metropolis filter~\cite{Michel2014JCP}
inverts the order of the product and the minimization and thus casts the
acceptance probability of a move into the same factorized form as the Boltzmann
weight:
\begin{equation}
    \pFact(\Conf \to \NewConf) = \prod_{M} \min \glc 1, 
\expb{-\beta \Delta
    U_{M}} \grc.
\label{equ:MetropolisFactorized}    
\end{equation}
The factorized filter in \eq{equ:MetropolisFactorized} and the Boltzmann
weight are now written as analogous products. Strictly speaking,  $M$ is a
generalized index denoting a factor ($\expb{-\beta \Delta U_M}$ or $ \min\glc
1, \expb{-\beta \Delta U_{M}} \grc$). It is for simplicity that we refer to
$M$ as a \quot{factor} rather than a \quot{generalized index for the Boltzmann
factor and the filter factor}.

The factorized Metropolis filter satisfies the detailed-balance condition:
\begin{equation}
\pi(\Conf)\pFact(\Conf\to \NewConf)  =
 \pi(\NewConf) \pFact(\NewConf \to \Conf).
\label{equ:DetailedBalance}
\end{equation}
This is evident if there is only a single factor
($U=U_M$ in \eq{equ:PotentialFactorized}  so that
\eqtwo{equ:MetropolisProduct}{equ:MetropolisFactorized} are identical), because
the Metropolis algorithm itself is well known to satisfy it:
\begin{equation}
\underbrace{\pi(\Conf) \pMet(\Conf \to \NewConf)}_{
\FCAL^{\text{Met}} _{\Conf \to \NewConf}
} = 
\underbrace{\pi(\NewConf) 
\pMet(\NewConf \to \Conf) }_{
\FCAL^{\text{Met}} _{\NewConf \to \Conf} }.
\label{equ:DetBalMet}
\end{equation}
If there is more than one factor, $\pFact $ also satisfies detailed balance
because the Boltzmann weight $\pi$ of \eq{equ:BoltzmannFactorized} and the
factorized Metropolis filter $\pFact$ of \eq{equ:MetropolisFactorized}
factorize (that is, break up) in exactly the same way and
\eq{equ:MetropolisFactorized}, on the level of a single factor, is again
equivalent to the Metropolis algorithm.

Applying the Metropolis filter $\pMet$ of \eq{equ:MetropolisProduct} is
equivalent to drawing a Boolean random variable:
\begin{equation}
    X^{\text{Met}}(\Conf \to \NewConf) = 
\begin{cases}
\text{\quot{True}} \quad &\text{if}\ \ranb{0,1} <
\pMet (\Conf \to \NewConf)
\\ 
\text{\quot{False}} \quad &\text{else} , \\
\end{cases}
\label{equ:MetropolisBoolean}                               
\end{equation}
where \quot{True} means that the move from configuration $\Conf$ to
configuration $\NewConf$ is accepted.  Similarly, the factorized Metropolis
filter $\pFact$ could be applied by drawing a single Boolean random variable
with $\pFact$ replacing $\pMet$ in \eq{equ:MetropolisBoolean}.  However,
because $\pFact \le \pMet$, this would yield a less efficient algorithm.
We rather view the factorized Metropolis filter as a conjunction of Boolean
random variables:
\begin{equation}
    X^{\text{Fact}}(\Conf \to \NewConf)  = \bigwedge_{M \in \SETOFM}
X_{M}(c_M \to c'_{M}).
\label{equ:MetropolisFactorizedBoolean}                               
\end{equation}
Now, $X^{\text{Fact}}(\Conf \to \NewConf)$ is \quot{True} if the independently
drawn factorwise Booleans $X_{M}$ are all \quot{True}:
\begin{equation}
X_{M} = 
\begin{cases}
\text{\quot{True}} \quad &\text{if}\ \ranb[M]{0,1} <\expa{-
\beta \Delta U_{M}} ,
\\ 
\text{\quot{False}} \quad &\text{else} , 
\end{cases}
\end{equation}
where the uniform random variables $\ranb[M]{0,1}$ are mutually independent for
all $M$.

The conjunction of  \eq{equ:MetropolisFactorizedBoolean} formulates the
consensus principle: In order to be accepted, the move $\Conf \to \NewConf$
must be independently accepted by all factors $M$.  For example, for a
homogeneous $N$-particle system with pair
factors $\FACTOR{\SET{i,j}}{\PAIRTYPE}$, the move of a single particle $k$ 
must be individually accepted by the factors $\FACTOR{\SET{k,j}}{\PAIRTYPE}\ 
\forall j \ne k$. In other words, the move of particle $k$ must be accepted by 
all other particles, each through its individual Metropolis filter.

For a continuously varying potential, the acceptance probability of a single
factor $M$ has the following infinitesimal limit:
\begin{multline}
\min \glc 1, \expb{-\beta \Delta U_M} \grc = 
\expb{-\beta \maxZeroa{\Delta U_M}}\\
\xrightarrow{\Delta U_M \to \diff U_M} 
1 - \beta \maxZeroa{\diff U_M},
\label{equ:AcceptanceFactors}
\end{multline}
where 
\begin{equation}
\maxZeroa{x} =  \max(0, x)
\label{equ:UnitRamp}
\end{equation}
is the unit ramp function of a real number $x$.  In this limit, the factorized
Metropolis filter becomes
\begin{equation} 
    \pFact(\Conf \to \NewConf) = 1 - \beta \sum_{M} 
\maxZeroc{\diff
    U_M (\Conf_M \to \NewConf_M) }  ,
\label{equ:MetropolisFactorizedInfinitesimal}    
\end{equation} 
and the total rejection probability for the move becomes a sum over factors:
\begin{equation}
1 - \pFact (\Conf \to \NewConf)  = \beta\sum_{M} \maxZeroc{\diff
    U_M (\Conf_M \to \NewConf_M)}.
\label{equ:MetropolisRejectionInfinitesimal}
\end{equation}
In ECMC, the infinitesimal limit generally corresponds to the continuous-time
displacement of a particle $k$ at position $\rvec_k = (x_k, y_k, z_k)$ and
it is usually along a coordinate axis. Supposing that this displacement is in
direction $\UnitX$, the differential of the factor potential becomes
\begin{equation}
 \diff U_M =  \FACDERIV{M}{k} \, \diff x_k ,
\end{equation}
where 
\begin{equation}
\FACDERIV{M}{k}(\SET{\rvec_i: i \in I_M}) = \partpart{U_M}{x_k},\quad (k 
\in I_M),
\end{equation}
is the factor derivative with respect to particle $k$. We then define the
factor event rate with respect to particle $k$ as
\begin{equation}
\EVRATE{M}{k} = \beta \maxZeroc{\FACDERIV{M}{k} } ,
\end{equation}
so that each of the terms $\maxZeroa{\diff U_M}$ becomes 
\begin{equation}
\beta \maxZeroa{\diff U_M} = \EVRATE{M}{k} \, \diff x_k .
\label{equ:DiffUEvRate}
\end{equation}
The event rate $\EVRATE{M}{k}$ yields the probability of an event being
triggered by particle $k$ within factor $M$. The total event rate
\begin{equation}
\TOTEVRATE_k(\SET{\rvec_1 \TO \rvec_N}) = 
\!\!\!\!\!\! \!\!\!\!\!\!
\sum_{M = \FACTOR{I_M}{T_M}: 
k \in I_M} 
\!\!\!\!\!\! \!\!\!\!\!\!
\EVRATE{M}{k}(\SET{\rvec_i: i \in I_M})
\label{equ:TotalEventRate}
\end{equation}
with respect to a particle $k$ naturally involves only event rates for factors
that contain  $k$ in their index set.

\subsection{Lifting and factorization schemes}
\label{sec:ECMCLifting}

The lifting concept~\cite{Diaconis2000} is central to ECMC. It lends
persistence to the individual Monte Carlo moves and thereby allows
one to take the zero-displacement limit. It is in this limit that the
sampling of factors becomes unique. We now describe the implementation of
a lifted irreversible Markov chain for the simulation of pair-interacting
particles~\cite{Michel2014JCP}, starting with a single pair.  We then
generalize\cite{Harland2017} the method to complex multi-particle potentials.

In a standard Markov-chain Monte Carlo algorithm, the rejection of a move of
some particle at time $s$ imposes that the state $\Conf(s+1)$  of the Markov
chain at time $s+1$  remains unchanged with respect to the state $\Conf(s)$ at
time $s$.  A new move is then proposed. For a local Monte Carlo algorithm in a
particle system, this new move normally consists in an independently sampled
displacement applied to another randomly chosen particle.  In order to
converge towards the correct stationary distribution $\pi$, we recall that the
Markov chain must satisfy the global-balance condition:
\begin{equation}
\FCAL_\Conf = \sum_{\OldConf} \FCAL_{\OldConf \to \Conf} = 
\sum_{\OldConf} \pi(\OldConf)p(\OldConf \to \Conf) = \pi(\Conf),
\label{equ:GlobalBalance}
\end{equation}
meaning that the total flow $\FCAL_c$ into a configuration $\Conf$
must equal its Boltzmann weight~\footnote{To simplify, we do not distinguish
between the filter, which is, strictly speaking, the acceptance probability of
a proposed move $\Conf \to \NewConf$, and the probability to move from $\Conf$
to $\NewConf$. In our context, the difference between the two is at most a
constant factor.}.
The detailed-balance condition of
\eq{equ:DetailedBalance} is only a special solution of \eq{equ:GlobalBalance}.
In addition to the global-balance condition, the Markov chain must
also be irreducible and aperiodic. These two conditions are easily
satisfied~\cite{Levin2008}; the former guarantees that any configuration
will eventually be visited, while the latter guarantees that the large-time
limit has no hidden periodicities.

\begin{figure}[htb]
\includegraphics{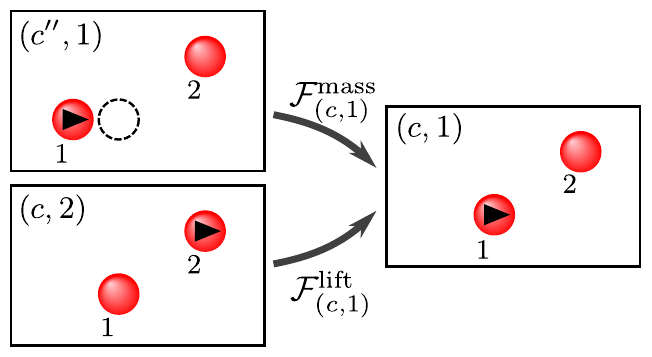} 
\caption{Mass flow (from \lifted{\OldConf}{\OneP}) and lifting flow (from
\lifted{\Conf}{\TwoP}) into a lifted configuration \lifted{\Conf}{\OneP},
corresponding to an accepted and a rejected particle move, respectively
(see \eq{equ:LiftedTwoParticles}). The total flow should equal the Boltzmann
weight $\pi(\Conf)$ in order to satisfy the global balance condition of
\eq{equ:GlobalBalance}.  }
\label{fig:DiscreteLifting}
\end{figure}

In ECMC, any physical configuration $c$ (that is, any set of particle
coordinates) is augmented (or \quot{lifted}\cite{Chen1999}) to include a
so-called lifting variable describing which particle is \quot{active}:
\begin{equation} 
\Conf \equiv \SET{\rvec_1 \TO
\rvec_N} \mapsto \lifted{\Conf}{\ActiveP}.  
\end{equation} 
In principle, the Boltzmann weight now depends on $\ActiveP$, but, for
simplicity, we require $\pi[\lifted{\Conf}{\ActiveP}] = \pi(\Conf)/N$ and
absorb the normalization factor $1/N$ into the zero of the potential and omit
it in the following.

In ECMC, furthermore, the particle \ActiveP\ (the active particle) remains
active for subsequent moves as long as they are accepted, and the displacement
(in the case that we will treat) is always the same \footnote{Strictly
speaking, the characteristics of the displacement are also to be included
among the lifting variables.}. For simplicity of notation, in the following,
the displacement $\Deltax$ is applied in the $\UnitX$ direction for all
moves so that the position $\rvec_\ActiveP $ is updated to $\rvec_\ActiveP +
\Deltax \UnitX$ for accepted moves. When a displacement $\rvec_\ActiveP  \to
\rvec_\ActiveP + \Deltax \UnitX$ is rejected by a target particle \TargetP, the
state of the lifted Markov chain changes in the augmented space as
\begin{equation}
\lifted{\Conf}{\ActiveP} \to \lifted{\Conf}{\TargetP},
\end{equation}
but the physical configuration $\Conf$ remains unchanged. Liftings
thus replace rejections.  The global-balance condition must be written
in terms of the augmented configurations, and the probability flow
$\FCAL_{\lifted{\Conf}{\ActiveP}}$ into each lifted configuration
$\lifted{\Conf}{\ActiveP}$ is then given by the sum of the \MassFlow\
flow $\FCAL^{\MassFlow}_{\lifted{\Conf}{\ActiveP}}$, that is,
flow corresponding to a particle displacement, and the lifting flow
$\FCAL^{\text{lift}}_{\lifted{\Conf}{\ActiveP}}$. This sum must equal the
statistical weight of $\lifted{\Conf}{\ActiveP}$ that, as discussed, equals
$\pi(\Conf)$:
\begin{equation}
 \FCAL_{\lifted{\Conf}{\ActiveP}} = 
 \FCAL^{\MassFlow} _{\lifted{\Conf}{\ActiveP}} + 
 \FCAL^{\text{lift}}_{\lifted{\Conf}{\ActiveP}} = \pi(\Conf). 
\end{equation}
In order to assure irreducibility of the Markov chain, one may change the
direction of motion, most simply by selecting from the set $\SET{\UnitX,
\UnitY, \UnitZ}$ in a way that does not need to be random (see the discussion
in \sect{sec:WaterIntrinsicRotations}). In ECMC, the process in between two
changes of direction is the eponymous \quot{event chain}. The length $\ell$ of
an event chain (the cumulative sum of the displacements), and the distribution
of $\ell$ are essential parameters for the performance of the algorithm.

To demonstrate that ECMC satisfies the global balance condition, and to study
the conditions on the lifting probabilities, we first consider a system of
two particles $\SET{\OneP,\TwoP}$. We may suppose, without restriction,
that the active particle is \OneP\, so that, at a given time, the lifted
configuration is $\lifted{\Conf}{\OneP}$. This lifted configuration can
only be reached from two other lifted configurations, one that differs
in the configuration variable, and the other in the lifting variable (see
\fig{fig:DiscreteLifting}). The lifted configurations and the corresponding
flows are:
\begin{equation}
  \begin{tikzcd}
   \lifted{\OldConf}{\OneP} 
   \arrow{dr}{\text{$\FCAL^{\MassFlow}_{\lifted{\Conf}{\OneP}}$ (\MassFlow\
flow)} }
   & 
\\
     & 
  \lifted{\Conf}{\OneP} 
     \\
    \lifted{\Conf}{\TwoP} 
\arrow[swap]{ur}{\text{$\FCAL^{\text{lift}}_{\lifted{\Conf}{\OneP}}$ (lifting 
flow)} } 
   \arrow{dr}{\text{$\FCAL^{\MassFlow}_{\lifted{\NewConf}{\TwoP}}$ 
 } }
   & 
\\
     & 
  \lifted{\NewConf}{\TwoP} 
\end{tikzcd}.
\label{equ:LiftedTwoParticles}  
\end{equation}
where 
$\Conf =  \SET{\rvec_\OneP, \rvec_\TwoP}$, 
$\OldConf = \SET{\rvec_\OneP -\Deltax \UnitX, \rvec_\TwoP}$, and
$\NewConf =  \SET{\rvec_\OneP, \rvec_\TwoP +\Deltax \UnitX } $.
The \MassFlow\ flow of the lifted algorithm from \lifted{\OldConf}{\OneP}\ to
\lifted{\Conf}{\OneP}\ equals the total Metropolis flow 
from the nonlifted configuration \OldConf\ to \Conf. Because of detailed
balance, the latter equals the (nonlifted)
Metropolis flow from \Conf\ to \OldConf, so that:
\begin{equation}
\FCAL^{\MassFlow}_{\lifted{c}{\OneP}} = 
\underbrace{\FCAL^{\text{Met}}_
{ \OldConf \to  \Conf } = 
\FCAL^{\text{Met}}_
{ \Conf \to  \OldConf } }_{\text{see \eq{equ:DetBalMet} }}
 = \pi(\Conf)  \pMet( \Conf \to \OldConf).
\label{equ:MovingFactorTwo}
\end{equation}
The lifting flow in \eq{equ:LiftedTwoParticles} equals  the rejection
probability of the Metropolis move $\Conf \to \NewConf$.  Because of
translational invariance ($\OldConf$ is a translated version of $\NewConf$), it
agrees with the Metropolis rejection probability of the move back from $\Conf$
to $\OldConf$:
\begin{multline}
\FCAL^{\text{lift}}_{\lifted{c}\OneP} = 
\pi(\Conf) \glc 1 - \pMet(\Conf \to \NewConf) \grc \\
=
\pi(\Conf) \glc 1 - \pMet(\Conf \to \OldConf) \grc.
\label{equ:MovingFactorThree}
\end{multline}
$\FCAL^{\MassFlow}_{\lifted{\Conf}{\OneP}}$ and
$\FCAL^{\text{lift}}_{\lifted{\Conf}{\OneP}}$ 
thus add up to the Boltzmann weight $\pi(\Conf)$, and global balance is
satisfied.  The validity of the lifted algorithm (which only satisfies global
balance, but breaks detailed balance) hinges on the fact that the underlying
Metropolis algorithm satisfies detailed balance and on the translation
invariance of the system.

\begin{figure*}[t]
\includegraphics{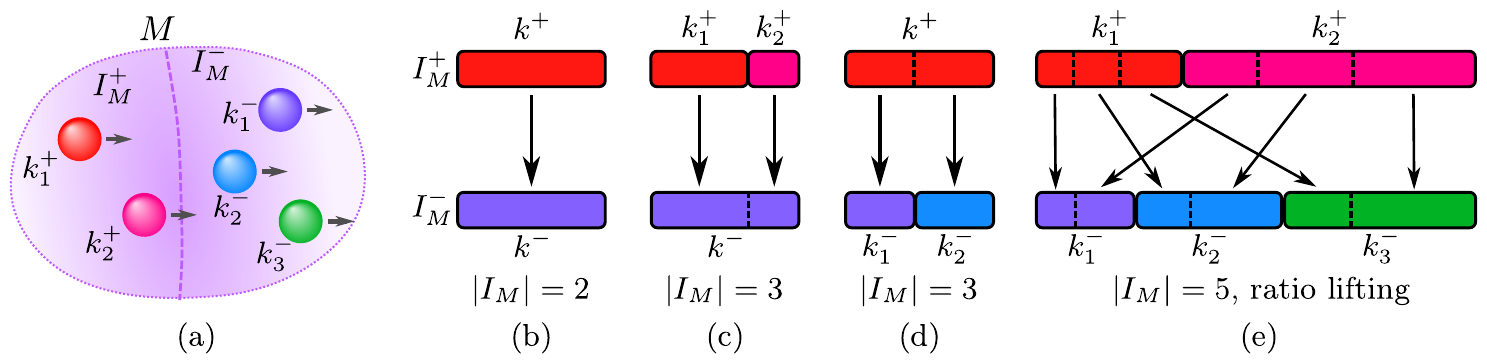}
\caption{Factors and lifting schemes. \subcap{a}: A factor $M$
consisting of $|I_M| = 5$ particles, split into non-empty sets $I_M^+$
(particles that increase the factor potential) and $I_M^-$ (see
\eqtwo{equ:MPlusDef}{equ:MMinusDef}).  \subcap{b--e}: Lifting schemes.
Unit branching $\gamma_{\UpP \to \DownP}=1$ and $\gamma_{\UpP_1 \to \DownP}=1,
\gamma_{\UpP_2 \to \DownP}=1$ for a pair-particle factor (\subcap{b}) and
a three-partice factor with $|I_M^-|=1$ (\subcap{c}), and \quot{ratio}
lifting scheme for $|I_M| = 3, |I_M^+| =1$ (\subcap{d}) and for $|I_M| > 3$
(\subcap{e}, see \eq{equ:RatioLiftingProbability}).  }
\label{fig:FactorBasics}
\end{figure*}

In the infinitesimal limit, for $N$ particles and a particle-pair
factorized potential, the total probability flow into a lifted configuration
\lifted{c}{\ActiveP} has up to $N$ components, namely $N-1$ lifting flows
from \lifted{c}{k} to \lifted{c}{\ActiveP}\ for $k \ne a$ and one \MassFlow\
move from \lifted{c'}\ActiveP{} to \lifted{c}\ActiveP, where $c'$ is again the
nonlifted configuration with $x_a$ replaced by $x_a - \diff x$. This corresponds
to one lifting flow $\FCAL^{\text{lift}}(k \to a)$ equivalent to that in
\eq{equ:LiftedTwoParticles} per target particle $k \ne a$, and a \MassFlow\
flow that is the infinitesimal analogue of that in \eq{equ:LiftedTwoParticles}.
Furthermore, a particle-pair potential may be further factorized according to
multiple factor types $T_M$; there then exist $N - 1$ lifting flows for each
factor $M$ consisting of two particles ($|I_M| = 2$, with $I_M$ the index set
of $M$). Of course, factors that do not contain $a$ in their index set do not
contribute to this flow.

Factors $M$ with more than two particles ($|I_M| > 2$) can also be handled
within the lifting framework~\cite{Harland2017} because, by translational
invariance, the sum over the factor derivatives with respect to particle $k$
satisfies:
\begin{equation}
\sum_{k \in I_M} \partpartshort{U_M}{x_k}(\SET{\rvec_i: i \in I_M}) = 0.
\label{equ:DerivativesBalance}
\end{equation}
It is useful to separate the particle indices  $k \in I_M$ of a factor
$M$ into two sets $I_M^+$ (with positive factor derivatives) and $I_M^-$
(negative factor derivatives) such that:
\begin{align}
     \UpP \in I_M^+ \Leftrightarrow\quad &   \partpartshort{U_M}{x_{\UpP}} > 0
     \label{equ:MPlusDef} \\
     \DownP \in I_M^- \Leftrightarrow\quad &   \partpartshort{U_M}{x_{\DownP}} 
< 
0, 
     \label{equ:MMinusDef}
\end{align}
where the factor derivatives satisfy
\begin{equation}
\sum_{\UpP \in I_M^+ } \partpartshort{U_M }{x_{\UpP}} = - \sum_{\DownP \in 
I_M^-} 
\partpartshort{U_M }{x_{\DownP}}
\label{equ:FactorEquality}
\end{equation}
(see \subfig{fig:FactorBasics}{a}).

The \MassFlow\ flow into a lifted configuration \lifted{\Conf}{\UpP} with  
$\UpP \in I_M^+$ by itself satisfies global balance,
\begin{multline}
\FCAL^{\MassFlow}_{\lifted{\Conf}{\UpP}} = 
\pi_{M}(\OldConf) \pMet_M(\OldConf \to \Conf)   \\ = \pi_{M}(\Conf) 
\pMet_M(\Conf 
 \to
\OldConf)  = \pi_M(\Conf), 
\end{multline}
so that there can be no additional lifting moves into \lifted{\Conf}{\UpP}.
This implies that lifting moves are always of the type $\lifted{c}{\UpP} \to
\lifted{c}{\DownP}$, that is, from an active particle in  $I_M^+$ to a target
particle in  $I_M^-$.  In contrast, the \MassFlow\ flow into the configuration
\lifted{\Conf}{\DownP}\ is smaller than $\pi_M(c)$:
\begin{multline}
\FCAL^{\MassFlow}_{\lifted{\Conf}{\DownP}} =  
\pi_{M}(\OldConf) \pMet_M(\OldConf \to \Conf)  \\ 
= \pi_{M}(\Conf) \pMet_M(\Conf  \to \OldConf) \\
= \pi_M(\Conf) ( 1 + \underbrace{\beta \partpartshort{U_M}{x_{\DownP}}}_{<0\, 
\text{(see \eq{equ:MMinusDef})\!\!\!\!\!\!\!\!\!\!\!\!\!\!\!\!\!\!}} \diff x ). 
\label{equ:PhysicalLargeFactor}
\end{multline}
The total lifting
flow into \lifted{\Conf}{\DownP} comes from all lifted configurations
\lifted{\Conf}{\UpP} with  $\UpP \in I_M^+$:
\begin{equation}
\FCAL^{\text{lift}}_{\lifted{\Conf}{\DownP}} = \pi_M(\Conf) \beta \sum_{\UpP 
\in 
I_M^+} 
\partpartshort{U_M}{x_{\UpP}} \diff x \gamma_{\UpP \to \DownP},
\label{equ:LiftingLargeFactor}
\end{equation}
where $\gamma_{\UpP \to \DownP}$ is the lifting probability from \UpP\ to
\DownP\ once the displacement of \UpP\ has been rejected.  In order for global
balance to hold, \eqtwo{equ:PhysicalLargeFactor}{equ:LiftingLargeFactor}
must add up to $\pi(\Conf)$ for all $\DownP \in M^-$. Therefore, and for the
algorithm to be rejection-free, one needs~\cite{Harland2017}:
\begin{align}
 \forall \DownP \in  I_M^-:  \underbrace{\partpartshort{U_M}{x_{\DownP}}}_{< 0} 
+ \!\!
\sum_{\UpP 
\in I_M^+} \underbrace{\partpartshort{U_M}{x_{\UpP}}}_{> 0} \gamma_{\UpP \to 
\DownP} 
&= 0, 
\label{equ:HarlandLiftings}
\\
 \forall \UpP \in  I_M^+:  \sum_{\DownP \in I_M^-} \gamma_{\UpP \to \DownP} &= 
1.
\label{equ:SumOfLiftings}
\end{align}

\Eqtwo{equ:HarlandLiftings}{equ:SumOfLiftings} can be visualized
as $|I_M^+|$ intervals of length $\partpartshort{U_M}{x_{\UpP}}$
placed on the upper row of a two-row table, and of $|I_M^-|$ intervals
of length $|\partpartshort{U_M}{x_{\DownP}}|$ on the lower row (see
\subfig{fig:FactorBasics}{b-e}). The total lengths of the two rows are equal
(see \eq{equ:FactorEquality}), and $\gamma_{\UpP \to \DownP} $ is the fraction
of the interval $\UpP$ on the upper row that lifts into $\DownP$ on the lower
row.  \Eq{equ:HarlandLiftings} describes a conservation of the interval lengths
from the upper row to the lower row.

For a pair factor ($| I_M | =2$), each row has one element, and
the lifting is unique ($\gamma = \gamma_{\UpP \to \DownP}= 1$, see
\subfig{fig:FactorBasics}{b}).  For a three-particle factor ($|I_M| =3$), if $|
I_M^+ | =2$, again clearly $\gamma_{\UpP \to \DownP} = 1$ for each one of the
particles $\UpP \in I_M^+$ (see \subfig{fig:FactorBasics}{c}).  If $|I_M^+| =
1$ and $|I_M^-| =2$, then \eq{equ:HarlandLiftings} yields the unique branching
probabilities~\cite{Harland2017} from \ActiveP\ to $\DownP_1$ and $\DownP_2$:
\begin{align}
\gamma_{\UpP \to \DownP_1} &
= - \frac{\partpartshort{U_M}{x_{\DownP_1}} } 
{\partpartshort{U_M}{x_{\UpP}}}
\propto | \partpartshort{U_M}{x_{\DownP_1}}| , 
 \\
\gamma_{\UpP \to \DownP_2} &
= - \frac{\partpartshort{U_M}{x_{\DownP_2}} } {\partpartshort{U_M}{x_\UpP}}
\propto | \partpartshort{U_M}{x_{\DownP_2}}|, 
\end{align}
which is readily understood from \subfig{fig:FactorBasics}{d}.  Analogously,
for factors with $|I_M| > 3$, the \quot{ratio} lifting corresponds to
cutting up each element in the upper row of the table into pieces of length
proportional to the elements in the lower row so that
\begin{equation}
\gamma_{\UpP \to \DownP} = 
\frac {\left | \partpartshort{U_M}{x_{\DownP}} \right |  } {\sum_{k^- \in 
I_M^-} 
\left | \partpartshort{U_M}{x_{k^-}} \right | }
\label{equ:RatioLiftingProbability}
\end{equation}
(see \subfig{fig:FactorBasics}{e}).  For factors with more than three particles
($|I_M| > 3$), the \quot{ratio} lifting scheme is not 
unique~\cite{Harland2017}. We
will make use of this freedom, in \secttwo{sec:Dipoles}{sec:Water}, for factors
with up to six particles corresponding to the atoms of two \hoh\ molecules.

\subsection{Event-driven and cell-veto methods}
\label{sec:EventDrivenCellVeto}

The implementation of ECMC differs notably from that of the Metropolis
algorithm,  both because of the continuous-time nature of the Markov chain,
which can be simulated without approximations using the event-driven
approach~\cite{Peters2012}, and because of the consensus property, which
can be checked in \bigO{1} operations via the cell-veto method, even
for infinite-ranged interactions~\cite{KapferKrauth2016}. It is these two
features that we explore in the present section. The intent is to overcome the
limitations of time-driven ECMC which considers a finite move $\Deltax \UnitX$
of the active particle:
\begin{equation}
\SET{\rvec_1 \TO \rvec_\ActiveP \TO \rvec_N} \to 
\SET{\rvec_1 \TO \rvec_\ActiveP + \eta \UnitX \TO \rvec_N}.
\label{equ:TimeDrivenMove}
\end{equation}
This move is either accepted (and then repeated) or it leads to a rejection
(by a factor $M \in \SETOFM$ containing particle $\ActiveP$), and it gives
rise to a  lifting (or possibly to multiple simultaneous liftings).  The 
complexity
of time-driven ECMC is \bigO{|\SET{M: \ActiveP \in I_M}| } per displacement
$\Deltax \UnitX$. Time-driven ECMC has a discretization error, as it becomes
inconsistent if more than one factor simultaneously rejects the move in
\eq{equ:TimeDrivenMove}. The parameter $\Deltax$ must be small enough for
multiple rejections to be rare. Time-driven ECMC is thus slow, especially for
long-ranged interactions, and inexact. It is useful only for testing.

The finite-move ECMC can be implemented as an event-driven, rather than as a
time-driven, algorithm~\cite{Peters2012,Bortz1975}, and because all factors are
independent, we may consider a single one of them.  In the above time-driven
ECMC, if the move in \eq{equ:TimeDrivenMove} (the first move, $m=1$) is
accepted, another displacement of magnitude $\Deltax$ is attempted. The $l$th 
move is:
\begin{multline}
\SET{\rvec_1 \TO \rvec_\ActiveP + (l-1) \Deltax \UnitX \TO \rvec_N} \\
\to \SET{\rvec_1 \TO \rvec_\ActiveP + l \Deltax \UnitX \TO \rvec_N}.
\label{equ:EventDrivenMove}
\end{multline}
After $m-1$ acceptances, finally, the $m$th such move is rejected (and leads to
a lifting).  The parameter $m$ is itself a random variable distributed with a
factor-dependent probability
\begin{equation}
 p_M(m) = 
 \underbrace{
 \prod_{l=1}^{m-1}
 \expa{-\beta \maxZeroa{\Delta U_{M}}(l)} 
}_{\text{accepted; see \eq{equ:AcceptanceFactors}}}
 \overbrace{
 \glc 1-\expa{-\beta \maxZeroa{\Delta U_M}(m) } \grc}^{\text{move $m$ 
rejected}} ,
\label{equ:EventProb}
\end{equation}
where $\maxZeroa{\Delta U_{M} }(l)$ is the change $\maxZeroa{\Delta U_M}$
corresponding to the $l$th move in \eq{equ:EventDrivenMove}.  The variable
$m$ can be sampled from \eq{equ:EventProb}, and the move $\rvec_\ActiveP \to
\rvec_\ActiveP + (m-1) \Deltax \UnitX$ accepted in one step.  Although the
right-hand side of \eq{equ:EventProb}, gives a probability distribution
for the displacement of the active particle \ActiveP, it only depends on
the positive increments of the factor potential.  In the continuum limit
$\Deltax \to 0$, the second term on the right-hand side becomes $\beta \diff
\maxZeroa{U_M}(\rvec_\ActiveP + \Deltax_M \UnitX) $, that is, the factor event
rate of \eq{equ:DiffUEvRate}, where $\Deltax_M$ is the total displacement 
before a rejection by factor $M$ takes place. In this limit, the exponent in 
the first term on the right-hand
side contains the integral of the factor event rate for the displacement of
$\rvec_\ActiveP$ between $0$ and $\eta_M$. This gives the 
probability density~\cite{Peters2012}:
\begin{equation}
p_M(\maxZeroa{U_M}) = \beta \expb{-\beta  \maxZeroa{U_M}}.
\label{equ:EventDrivenECMC}
\end{equation}
In \eq{equ:EventDrivenECMC}, the exponential distribution is sampled by:
\begin{equation}
\beta \maxZeroa{U_M} = -\logd{\ranb[M]{0,1}},
\label{equ:SampleEta} 
\end{equation}
\begin{figure}[htb]
\includegraphics{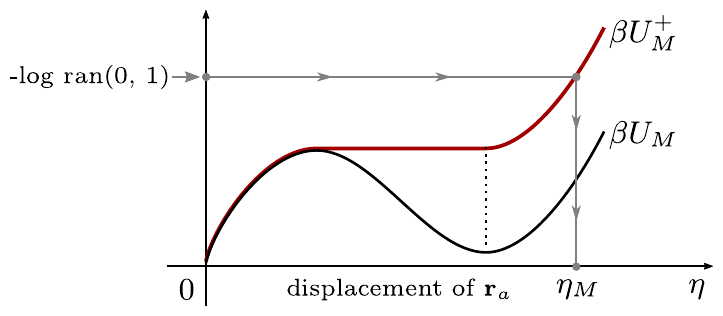}
\caption{Event-driven ECMC\cite{Peters2012} for a two-particle factor $M$.  The
integral of the factor derivative multiplied with $\beta$ equals $\beta
U_M$, whereas the integral of the event rate (\emph{in red}) must equal $\beta
\maxZeroa{U_M}$, which is sampled from \eq{equ:SampleEta}.  The calculation of
the displacement $\Deltax_M$ from the sampled value of $\beta \maxZeroa{U_M} =
-\loga{ \ranb{0,1}}$ is indicated by arrows.}
\label{fig:EventDrivenECMC}
\end{figure}
where
\begin{multline}
  \underbrace{\beta \maxZeroa{U_M}(\rvec_\ActiveP + \Deltax_M \UnitX) 
}_{\text{sampled via \eq{equ:SampleEta}}}  \\
 =  \int_{0}^{\Deltax_M} 
   \underbrace{\beta \maxZeroc{\partpartshort{U_M}{x_a}
   (\SET{\rvec_\ActiveP + \Deltax \UnitX, \rvec_k: k \in I_M} )}
   }_{\text{factor event rate, see \eq{equ:DiffUEvRate}}}
    \diff \Deltax .
\label{equ:CumulativeEventRate}   
\end{multline}
In other words, $\beta \maxZeroa{U_M} $ is the cumulative event rate of
\eq{equ:DiffUEvRate}. \Eq{equ:CumulativeEventRate} is an implicit relation for
the limiting displacement $\Deltax_M$ at which the rejection takes place as a
function of the sampled value of $\beta \maxZeroa{U_M}$.  For a two-particle
factor $M = \FACTOR{\SET{\ActiveP,k}}{\PAIRTYPE}$, the integration of the pair 
event rate in  \eq{equ:CumulativeEventRate} consists in the replacement of the
potential $U_M$ by a related potential which is zero at $\rvec_\ActiveP$,
and where all the negative increments are replaced by horizontal lines (see
\fig{fig:EventDrivenECMC}).

As mentioned, the factors are independent, and each concerned factor $M$ 
provides a value $\Deltax_M$. The next event takes place at
\begin{equation}
   \Deltax = \min_{M: \ActiveP \in I_M} \Deltax_M
\end{equation}
and the factor which realizes this minimum (that is, $\Deltax$)
\begin{equation}
   \text{argmin}_{M: \ActiveP \in I_M} \Deltax_M
\end{equation}
is the one in which the lifting takes place.  For a continuous potential,
this factor is uniquely defined, and possible simultaneous events, due to
finite-precision arithmetic, are too rare to play a role.

The integration of the factor event rate in \eq{equ:CumulativeEventRate} can be
tedious if it cannot be cast into an explicit analytical form.  This will for
example be the case for the Coulomb potential in the merged-image framework
of \sect{sec:CoulombDerivatives}. In addition, the inversion of the factor
potential (the computation of $\Deltax_M$ in \eq{equ:CumulativeEventRate})
can be non-trivial.  Finally, this calculation must in principle be redone
for all the factors that contain the active particle $\ActiveP$. For a
long-ranged potential, this requires  $\bigO{N} = \bigO{| \SET{M \in \SETOFM:
i \in I_M}|}$  event-rate integrations and inversions per event.  The cell-veto
algorithm~\cite{KapferKrauth2016}, by use of a comparison function, avoids the
integration and the inversion of the event rate, and it moreover reduces the
overall complexity of ECMC to \bigO{1} per event.

\begin{figure}[htb]
\includegraphics{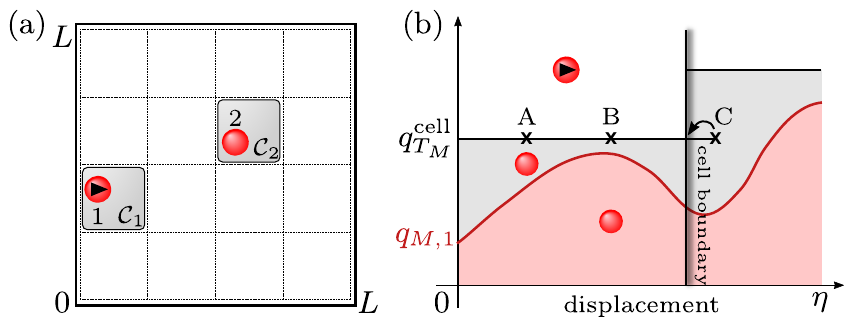}
\caption{Cell-veto algorithm for a two-particle factor  $M$.
\subcap{a}: Active particle \OneP\ in cell $\Cell_\OneP$ and
target particle \TwoP\, in cell $\Cell_\TwoP$. \subcap{b}: The event-rate
$\EVRATE{M}{\OneP}(\rvec_1,\rvec_2)$ is bounded from above by the cell-event 
rate
$\qCell_{T_M}(\Cell_\OneP, \Cell_\TwoP)$, which can be sampled trivially. A
cell event may either be rejected (at point \quot{A}) or confirmed (at point
\quot{B}) as a particle event (see \eq{equ:CellConfirmation}), while a cell
event taking place outside $\Cell_\OneP$ (at point \quot{C}) means that the
active particle $\OneP$  will be advanced towards the cell boundary.}
\label{fig:CellEventScheme}
\end{figure}

We again first consider a pair factor $\FACTOR{\SET{\OneP,\TwoP}}{\PAIRTYPE}$, 
with \OneP\ the active particle. The lifted position is \lifted{\Conf}{\OneP}\
(with $\Conf = (\rvec_\OneP, \rvec_\TwoP)$) and the displacement is again in
direction $\UnitX$ (as in the situation in \fig{fig:DiscreteLifting}). We
embed the two particles in disjoint cells $\Cell_\OneP$ and $\Cell_\TwoP$
(see \fig{fig:CellEventScheme}). The potentials that we consider here
are singular only at $ \rvec_{\OneP} = \rvec_{\TwoP} $, so that the event
rate for factor $M$ may be bounded by a constant \quot{cell-event} rate
$\qCell_{T_M}(\Cell_\OneP, \Cell_\TwoP)$:
\begin{equation}
\EVRATE{M}{\OneP}(
\rvec_{\OneP},\rvec_{\TwoP}
) \le \qCell_{T_M}(\Cell_\OneP, 
\Cell_\TwoP)\quad
\forall \rvec_\OneP \in \Cell_\OneP, \rvec_\TwoP \in \Cell_\TwoP,
\end{equation} 
where the right-hand side only depends on the factor type. This factor-type
dependence may take into account separate cell schemes that could for example
correspond to Coulomb interactions between isolated charges, dipole--dipole
interactions, or to the Lennard-Jones potential. (We recall that we do not 
differentiate the different Coulomb types for $2,4,6$ particles to ease 
notation.) In this work, the condition
$\Cell_\OneP \ne \Cell_\TwoP$ is adequate to ensure a reasonable value of the
cell-event rate. In other cases~\cite{KapferKrauth2016}, one must exclude
a local set of cells, and treat local neighbors outside the cell-veto
framework. Cell-event rates are easily tabulated in advance of the ECMC
computation proper.

The probability of the event taking place for an infinitesimal displacement 
$\diff x$ equals $q_{M, \OneP}(\rvec_{\OneP},\rvec_{\TwoP})\diff x$. Since
\begin{equation}
         \EVRATE{M}{\OneP}(\rvec_{\OneP}, \rvec_{\TwoP}) \diff x = 
         \underbrace{\qCell_{T_M}(\Cell_\OneP, \Cell_\TwoP) 
         \diff x}_{\text{infinitesimal}}  
        \underbrace{\frac{\EVRATE{M}{\OneP}(\rvec_{\OneP}, \rvec_{\TwoP})}
                   {\qCell_{T_M}(\Cell_\OneP, \Cell_\TwoP)}}_{\lesssim 1}, 
\label{equ:CellConfirmation}         
\end{equation}
the event can initially be sampled as a \quot{cell event} with
the constant infinitesimal probability $\qCell_{T_M}(\Cell_\OneP,
\Cell_\TwoP) \diff x $, before being confirmed with the finite probability
$\EVRATE{M}{\OneP}(\rvec_{\OneP},\rvec_{\TwoP}) / \qCell_{T_M}(\Cell_\OneP,
\Cell_\TwoP)\le 1$.  We may suppose that the cell event takes place at a lifted
configuration \lifted{\NewConf}{\OneP} with
\begin{align}
\NewConf &= (\rvec_\OneP + \eta \UnitX, \rvec_\TwoP) \label{equ:CPrime}\\
  \pi(\eta) & = \expc{- \eta \qCell_{T_M}(\Cell_\OneP, \Cell_\TwoP)}
              , \label{equ:CellEventExpDistr}
\intertext{where $\eta$ can be sampled via} 
  \eta & = -\logc{\ranb{0,1}}/\qCell_{T_M}(\Cell_\OneP, \Cell_\TwoP).
\end{align}

Three outcomes are possible for the sampled values of $\Deltax$ and the
subsequent confirmation step.  First, the cell event may correspond to
a configuration $\NewConf$ (in \eq{equ:CPrime}) that is already outside
the active-particle cell ($\NewConf \not\in \Cell_\OneP$).  In this case,
the move is $\lifted{\Conf}{1} \to \lifted{\OldConf}{1}$, where $\OldConf$
is the configuration intersecting the trajectory of particle \OneP\ with the
boundary of $\Cell_{\OneP}$.  Such a cell-boundary event moves the particle,
but does not trigger a lifting.  Second, the cell event may take place at
a configuration $\NewConf \in \Cell_\OneP$ but fail to be confirmed as an
event (because a uniform random number $\rand{0,\qCell_{T_M}(\Cell_\OneP,
\Cell_\TwoP)} > \EVRATE{M}{\OneP}(\rvec_{\OneP}, \rvec_{ \TwoP})$) (see
the second term on the right-hand side of \eq{equ:CellConfirmation}).
In this case, the move is $\lifted{\Conf}{1} \to \lifted{\NewConf}{1}$ and
no lifting takes place.  Third, a cell event may take place at a position
$\NewConf \in \Cell_\OneP$ and it is confirmed as an event.  This event
induces a lifting \lifting{\NewConf}{1}{2} (see \fig{fig:CellEventScheme}b).
In this whole process, the factor derivative $\FACDERIV{M}{1}$ is evaluated
only when a cell event is triggered from the exponential distribution in
\eq{equ:CellEventExpDistr}. The costly integration of the factor event rate in
\eq{equ:CumulativeEventRate} is thus avoided.

For an $N$-particle system, the cell-veto algorithm organizes the search
of the next lifting in \bigO{1} operations.  It suffices to choose
a regular grid of cells such that, normally, only a single particle
belongs to each cell.  (Exceptional double-cell occupancies can be handled
easily~\cite{KapferKrauth2016}.)  In this case, the total event rate with
respect to factor type $T_M$ for an active particle in \ActiveCell\ is bounded
by the total cell event rate
\begin{equation} 
Q^{\text{cell} }_{T_M}(\ActiveCell) = \sum_{\text{cells}\ \TargetCell \ne 
\ActiveCell}
\qCell_{T_M}(\ActiveCell, \TargetCell) .
\end{equation} 
In a translationally invariant system, the total cell event rate does not
depend on the active cell, so that $Q^{\text{cell} }_{T_M}(\ActiveCell)
\equiv Q^{\text{cell} }_{T_M}$, a constant that is computed before
the ECMC simulation starts from the total number of cells that scales
as \bigO{N}. The next cell event is obtained from an exponential
distribution with parameter $Q^{\text{cell}}_{T_M}(\ActiveCell)$. This
event corresponds to cell $\TargetCell$ with probability $\propto
\qCell_{T_M}(\ActiveCell, \TargetCell)$, posing a discrete
sampling problem that can be solved in \bigO{1} by Walker's
algorithm~\cite{Walker1977AnEfficientMethod,KapferKrauth2016}.

The cell-veto algorithm samples the Boltzmann distribution without performing
the event-rate integration in \eq{equ:CumulativeEventRate}. It requires  only
\bigO{1} factor-potential evaluations per event in an $N$-particle system. As a
consequence, the total potential of \eq{equ:PotentialFactorized} is not updated
and the potential remains unknown as the Markov chain evolves. This is what
sets ECMC apart from traditional simulation approaches.

\section{ECMC Coulomb algorithms}
\label{sec:Coulomb}

In a three-dimensional simulation box with periodic boundary conditions,
the Coulomb potential is only conditionally convergent for a charge-neutral
system, and it is infinite for a system with a net charge. Finiteness of the
potential can be recovered in both cases if each point charge is compensated by
a background charge distribution.  Traditionally, this is chosen as uniform
within the simulation box~\cite{deLeeuw1980-1}. The precise association
of each background charge with its point charge is not unique. This leads
to different electrostatic boundary conditions, which are linked to the
polarization state of the simulation box.  Consistency imposes a distinct
fluctuation theorem\cite{deLeeuw1980-1} for each choice of boundary condition
when computing macroscopic physical properties such as the dielectric
constant. Alternatively  to the uniform compensating background charge,
in ECMC, a line-charge model was introduced~\cite{KapferKrauth2016}. In
this model, the background charge distribution is one-dimensional and
the factor derivatives are absolutely convergent. The potential for
different variants of the line-charge model can be absolutely or conditionally
convergent.

\begin{figure}[htb]
\includegraphics{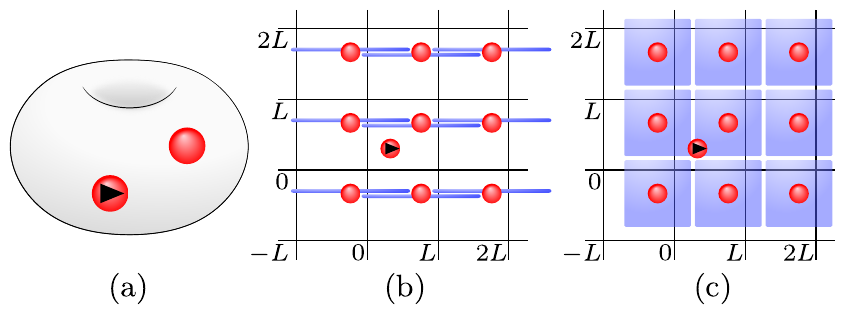}
\caption{Periodic two-particle Coulomb system. \subcap{a}: Toroidal
representation corresponding to a merged-image factor.  \subcap{b}: Line-charge
representation. The target point-charge particle and each of its copies are
compensated by line charges of length $2L$. The active particle inside the
central simulation box $[0,L)^3$ is not replicated.  \subcap{c}: Compensating
volume-charge representation corresponding to \quot{tin-foil} boundary
conditions.}
\label{fig:LineChargesAndGeometry}
\end{figure}

As discussed in \sect{sec:ECMCFactorized}, ECMC allows for different
Coulomb factor sets, that may influence the convergence properties of
the algorithm, although the steady state is invariably given by the
Boltzmann distribution.  Roughly, there are two inequivalent Coulomb
factorizations~\cite{KapferKrauth2016}.  Firstly, the periodic two-particle
problem can be embedded on a three-dimensional torus and the potential
merged from all the topologically inequivalent minimal paths between particles
(see \subfig{fig:LineChargesAndGeometry}{a}).  For two particles,
$\SET{\OneP, \TwoP}$, this \quot{merged-image} system has a single factor
$\FACTOR{\SET{\OneP, \TwoP}}{\TYPE{Coulomb}}$. For $N$ particles, this gives
the factor set
\begin{equation}
\SET{
\FACTOR{\SET{i, j}}{\TYPE{Coulomb}}: i< j \in \SET{1 \TO N} }.
\end{equation}
In general, the merged-image factors may comprise more than two particles,
but they do not distinguish between the different images of a local
configuration (for example an \hoh\ molecule).  Secondly, we may picture the
three-dimensional periodic system as an infinite number of periodic images
of the simulation box indexed by an integer vector $\nvec \in \ZZ^3$. For two
particles already, this \quot{separate-image} system  has an infinite number of
factors  and for $N$ particles, the factor set is
\begin{equation}
\SET{
\FACTOR{\SET{i, j}}{\TYPE[\nvec]{Coulomb}}: i< j \in \SET{1 \TO N}, \nvec \in 
\ZZ^3 }.
\label{equ:FactorizedCoulombSec2}
\end{equation}
More generally, an individual \quot{separate-image} factor may describe an
image of certain particles inside the simulation box.

The aim of this section is threefold. First, we present the tin-foil and
the line-charge Coulomb formulations and then demonstrate that, although
the potentials differ, the Coulomb factor derivatives (that for
pair factors yield the event rates) are identical. Second, we discuss two
efficient algorithms for the merged-image Coulomb derivatives of a pair of
particles, one algorithm from the tin-foil perspective and the other summing up
line-charge derivatives. Third, we set up an ECMC simulation for two particles
in a periodic three-dimensional simulation box in order to validate that the
merged-image and the separate-image factor sets indeed show indistinguishable
equilibrium properties. We then discuss possible applications for both
factorizations.

\subsection{Tin-foil electrostatics within ECMC}
\label{sec:CoulombTinFoils}

The traditional treatment of electrostatic interactions with periodic boundary
conditions is based~\cite{deLeeuw1980-1} on a large spherical aggregate of
images of the three-dimensional cubic simulation box. The polarization state
of the simulation box is expressed through electrostatic boundary conditions.
With \quot{tin-foil} boundary conditions, the potential
of $N$ particles $i \in \SET{1 \TO N}$ of charge $c_i$ (in units where the
Coulomb potential between two point charges in free space is $U_{ij}=c_i
c_j/|\rvec_{ij}|$), is~\cite{deLeeuw1980-1}:
\begin{multline}
U_{\COULOMB}(\SET{\rvec_1 \TO \rvec_N}, \SET{c_1 \TO c_N})  \\
= \frac{1}{2}\sum_ 
{i=1}^{N} c_i \psi(\rvec_i) 
+U_{\text {self}}(\alpha),
\label{equ:CoulombPotentialGeneral}
\end{multline}
with the electrostatic potential $\psi$:
\begin{align}
\psi(\rvec_i) = \, &
\sum_{j \ne i=1}^N c_j \glc \sum_{\nvec \in \ZZ ^3} \frac{{\rm erfc} (\alpha | 
\rvec_{ij}+\nvec L |)}{|
\rvec_{ij} + \nvec L |} \right.  \nonumber\\
& \left. + \frac{4 \pi }{L^3}\sum_{\qvec\ne (0,0,0)}
\frac{\expa{- \qvec^2 / (4\alpha^2)}}{\qvec^2}\cosb{\qvec\cdot \rvec_{ij}} \grc 
,
\label{equ:EwaldCoulFunction}
\end{align}
where the Fourier-space sum is over $\qvec=2 \pi \mvec/L$ with $\mvec \in \ZZ 
^3$. The self-energy contribution $U_{\text {self}}(\alpha)$ is
independent of the particle positions, and drops out of our considerations,
which are only concerned with derivatives of the potential.  The left-hand
side of \eq{equ:CoulombPotentialGeneral} is independent of the convergence
factor $\alpha > 0$, which however influences the speed of evaluation
of \eq{equ:EwaldCoulFunction}.  Direct evaluation of the sums for $N$
point charges leads to an optimal choice $\alpha \sim N^{1/6}/L $, and a
scaling in operations \bigO{N^{3/2}}. The Particle--Mesh Ewald method uses an
interpolating mesh to approximate the Fourier sum, leading to $\bigO{N\log{N}}$
operations to evaluate the potential. In merged-image ECMC we only use
\eq{equ:EwaldCoulFunction} for $N=2$ with $\alpha =\bigO{1/L}$, and evaluate
the derivative of the Coulomb potential to machine precision with \bigO{1}
effort.

We continue, as in \sect{sec:ECMCLifting}, with a two-particle factor
$\FACTOR{\SET{\OneP,\TwoP}}{\TYPE{\COULOMBTYPE}}$. The tin-foil factor 
derivative is given by:
\begin{multline}
  \FACDERIV{
  \FACTOR{\SET{\OneP,\TwoP}}{\TYPE{\COULOMBTYPE}} }{\OneP}
(\rvec_{\OneP \TwoP}, \SET{c_\OneP,c_\TwoP})  \\
=   \qtilde_{\text{Real}}(\rvec_{\OneP \TwoP}) +
  \qtilde_{\text{Four.}}(\rvec_{\OneP \TwoP}), 
\label{equ:CoulombDirectDeriv}
\end{multline}
with the real-space derivative $\qtilde_{\text{real}}$
\begin{align}
\qtilde_{\text{real}}(\rvec_{\OneP \TwoP}) = & 
    c_\OneP c_{\TwoP} \sum_{\nvec \in \ZZ ^3} \frac{\rvec_{\OneP \TwoP} + n_x 
L}{| 
  \rvec_{\OneP \TwoP} + \nvec L |^2} \left[
  \frac{{\rm erfc (\alpha | \rvec_{\OneP \TwoP}+\nvec L|) }}{|
  \rvec_{\OneP \TwoP}  + \nvec L |} \right. \nonumber\\
  & \left. + \frac{2\alpha \expa{-\alpha^2 |
  \rvec_{\OneP \TwoP}  + \nvec L
    |^2}}{\pi^{1/2}} \right], \label{equ:DirectDerivPos}
\end{align}
and the Fourier-space derivative $\qtilde_{\text{Four.}}$
\begin{align}
\qtilde_{\text{Four.}}(\rvec_{\OneP \TwoP}) = &  c_\OneP c_{\TwoP} \frac{4 
\pi}{L^3} \sum_{\qvec\ne 0}q_x\frac{\expa{- 
  \qvec^2 /(4 \alpha^2)}}{\qvec^2}\sinb{\qvec\cdot  \rvec_{\OneP \TwoP}}.
  \label{equ:DirectDerivFourier}
\end{align}
For two particles and, more generally, for pair factors in an
$N$-particle system, the merged-image Coulomb pair-event rate, from
\eq{equ:CoulombDirectDeriv}, is given by:
\begin{multline}
 \EVRATE{
  \FACTOR{\SET{\OneP,\TwoP}}{\TYPE{\COULOMBTYPE}} }{\OneP}
(\rvec_{\OneP \TwoP}, \SET{c_\OneP,c_\TwoP}) \\ =
\beta \maxZeroc{ \FACDERIV{
  \FACTOR{\SET{\OneP,\TwoP}}{\TYPE{\COULOMBTYPE}} }{\OneP}
(\rvec_{\OneP \TwoP}, \SET{c_\OneP,c_\TwoP})}.
\label{equ:metallicEventRate}
\end{multline}
In \secttwo{sec:Dipoles}{sec:Water}, we will consider dipole--dipole factors
with an index set comprising the four or six particles of two molecules
and the \quot{Coulomb} type corresponding to all the Coulomb interactions
between the two molecules.  The factor potential in this case is the sum
over Coulomb pairs within the factor, and the factor derivatives needed
in \eq{equ:HarlandLiftings} are the sum of a finite number of pairwise
Coulomb derivatives as in \eq{equ:CoulombDirectDeriv}. The evaluation of the
dipole--dipole factor derivatives remains of complexity \bigO{1} because
the number of elements in each factor remains finite as $N \to \infty$.
In ECMC, only a single factor has to be evaluated precisely for each move (see
\sect{sec:EventDrivenCellVeto}) whereas in traditional MCMC or MD computations
the Coulomb potential in \eq{equ:CoulombPotentialGeneral} or its derivatives
are computed for all $N$ particles.

\subsection{Line-charge model}
\label{sec:CoulombLineCharge}

In a large  periodically reproduced aggregate of the simulation box, the sum
over the Coulomb derivatives between a charged active particle and multiple
target images (without neutralizing backgrounds) is ill-defined. However,
the compensating uniform volume charge is not the only option to regularize
the sum, as the line-charge model~\cite{KapferKrauth2016}  and its
variants provide alternatives to tin-foil electrostatics. Here, straight
lines of charges are associated with each copy of the target particle,
and aligned with its direction of motion (in our example $\UnitX$, see
\subfig{fig:LineChargesAndGeometry}{b}).  Although the merged-image line-charge 
potential, in its simplest version, is itself not absolutely convergent, its 
factor
derivatives are unequivocally defined and equivalent to those obtained with
tin-foil boundary conditions.  By itself, the line charge neutralizes the
charge of the target particle, and (because it is centered) also creates an
object with zero dipole moment.  Previous work~\cite{KapferKrauth2016} used
line charges of  length $L$. Here, we consider lengths $pL$ with integer $p$
(see \subfig{fig:LineChargesAndGeometry}{b}).  The line charges are replicated
over a cubic lattice indexed by the lattice vector $\nvec$.  Lines of different
images meet (see \subfig{fig:LineChargesAndGeometry}{b}).  The Coulomb
potential of the line-charge model naturally differs from the one of the
tin-foil model because the background charge distributions are manifestly
different. However, the merged-image Coulomb derivative of the line-charge
model, relevant to ECMC, is identical to the tin-foil expression.

Explicitly, the contribution to the Coulomb derivative from an image $\nvec$
(with $\nvec = (0,0,0)$ the original simulation box) is
\begin{multline}
  \FACDERIV{\FACTOR{\SET{\OneP, \TwoP} }{\TYPE[\nvec]{\COULOMBTYPE} 
} }{1}(\rvec_{\OneP  \TwoP} ) 
   = c_\OneP c_{\TwoP} \gld \frac{\rvec_{\OneP \TwoP}  + n_x L}{| \rvec_{\OneP 
\TwoP}
    + \nvec L |^3} + \right. \\
   \left. \frac{1}{p L} \glc \frac{1}{| \rvec_{\OneP \TwoP} + 
    L \glb \nvec + p\UnitX \grb/2 | } - 
    \frac{1}{| \rvec_{\OneP \TwoP} +
    L \glb \nvec - p\UnitX \grb/2 | } \grc \grd 
    \label{equ:imagewiseLineChargeDirectDeriv}
  \end{multline}
The line charge generates an electrostatic potential  at large separations,
$\rvec=L\nvec$, which varies with  a quadrupolar form.  Thus, in any
given direction the Coulomb derivative decays as $1/|\rvec|^4$. For
this reason, the sum over the images of the Coulomb derivatives of
\eq{equ:imagewiseLineChargeDirectDeriv} converges absolutely.  The merged-image
Coulomb derivative, in the line-charge formulation, is thus
\begin{multline}
\overbrace{
\FACDERIV{
  \FACTOR{\SET{\OneP,\TwoP}}{\TYPE{\COULOMBTYPE}} }{\OneP}
(\rvec_{\OneP \TwoP}, \SET{c_\OneP,c_\TwoP})}^{\text{tin-foil expression, 
\eq{equ:CoulombDirectDeriv}}} \\ =
  \underbrace{\sum_{\nvec} 
  \FACDERIV{\FACTOR{\SET{\OneP, \TwoP} }{\TYPE[\nvec]{\COULOMBTYPE} 
} }{1}(\rvec_{\OneP  \TwoP} )}_{\text{sum over line charges, 
\eq{equ:imagewiseLineChargeDirectDeriv}}}.
  \label{equ:lineChargeRate}
\end{multline}
To show this, we first consider the target particle $\TwoP$ in the simulation
box and all its images to be surrounded by a cube of neutralizing charge of
volume $L^3$ centered on the particle \TwoP\ and its images. This volume-charge
model (see \subfig{fig:LineChargesAndGeometry}{c}) is closely connected
to the the line-charge model (see \subfig{fig:LineChargesAndGeometry}{b}).
Point charge and associated volume charge have vanishing charge, dipole and
quadrupole moments (whereas the line-charge model, in its simplest form,
has a finite quadrupole moment).  We now compare spherical (radius $R \gg
L$) and cubic aggregates (of side $2R$) of target images, and study the
electrostatic potential within the central simulation box. In this process, the
active particle is not replicated, and it remains within the simulation box.
Due to the vanishing quadrupole moment of the volume charges, the difference
in the electrostatic potential on the particle \OneP\ in the spherical
and cubic aggregates decreases at least as fast as $1/R^2$. However the
electrostatic potential in the center of the spherical aggregate corresponds
to a zero-polarization state which is identical to the tin-foil expression of
\eq{equ:EwaldCoulFunction}.
\begin{figure*}[ht]
\includegraphics[width = \linewidth]{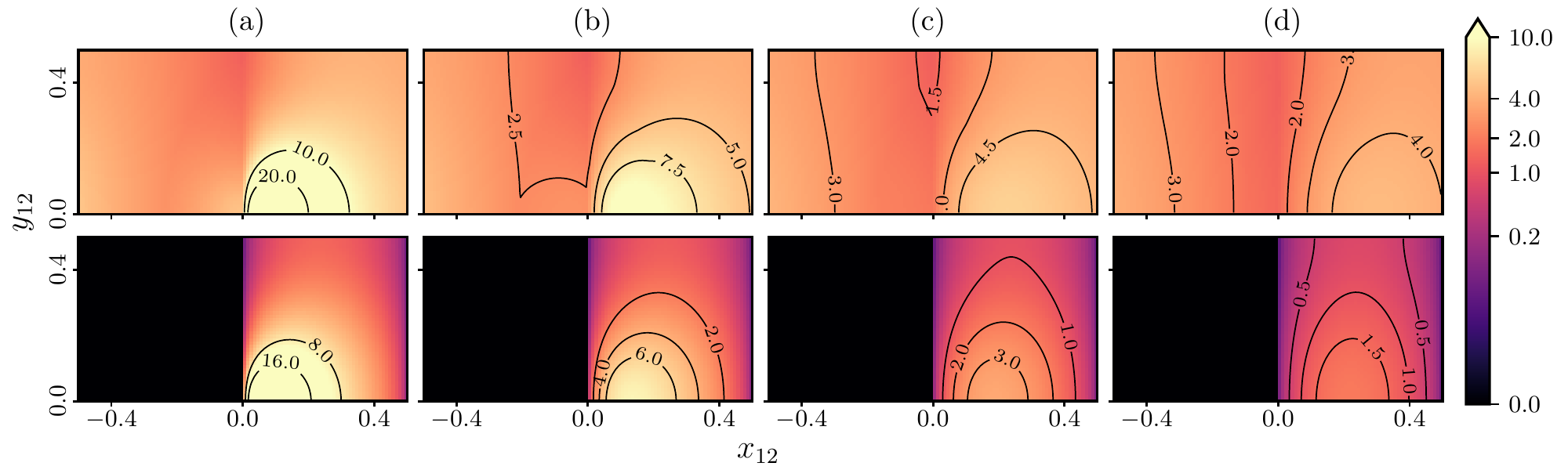}
\caption{Comparison of the sum of the separate-image event rates
$ \sum_\nvec \beta \maxZeroc{
\FACDERIV{
  \FACTOR{\SET{\OneP,\TwoP}}{\TYPE[\nvec]{\COULOMBTYPE}} }{\OneP}
(\rvec_{\OneP \TwoP}   )}$ (upper row)
and the merged-image event rate 
$ \beta \maxZeroc{
\FACDERIV{
  \FACTOR{\SET{\OneP,\TwoP}}{\TYPE{\COULOMBTYPE}} }{\OneP}
(\rvec_{\OneP \TwoP}   )}$ (lower row). In all panels $
\rvec_{\OneP\TwoP} = (x_{\OneP\TwoP}, y_{\OneP\TwoP}, z_{\OneP\TwoP})$
with 
\subcap{a}: $z_{\OneP \TwoP} = 0.1$,
\subcap{b}: $z_{\OneP \TwoP} = 0.2$,
\subcap{c}: $z_{\OneP \TwoP} = 0.3$, and
\subcap{d}: $z_{\OneP \TwoP} = 0.4$.
$L = 1$ and  $\beta c_\OneP c_{\TwoP} = 1$ throughout.}
\label{fig:PairEventPlot}
\end{figure*}

We now find explicit integral expression for the Coulomb derivative of an
aggregate of line charges and volume charges and show that the difference is
zero in the limit of  a large assembly.  We again consider the interaction
between an active particle and the cubic aggregate of the ${(2 \KK+1)}^3$
copies of the target particle (the central simulation box and its images). 
(The active particle is placed inside the simulation box.)  The Coulomb 
potential
between the active particle and a single target particle is
\begin{equation}
U_{\OneP \TwoP} = 4 \pi c_\OneP \int_{-\infty}^{\infty} 
          \frac{\diff^3 \qvec}{{(2\pi)}^3}\, e^{i {\qvec} \cdot \rvec_{\OneP 
\TwoP} }
\frac{\rho_\TwoP({\qvec})}{|\qvec|^2}, 
\label{equ:Uat}
\end{equation}
where $\rho_\TwoP({\qvec})$ is the  structure factor of the target particle
and the background. We now sum over the images, separated by a multiple of the
simulation box size $L$ along each axis. This requires evaluating the sum
\begin{equation}
D_K(q_x) = \sum_{l=-K}^K  \expa{i q_x l L} = \frac{\sinc{q_x L (K+1/2) } }{\sin
  (q_x L/2) }, 
\label{equ:kernel}
\end{equation}
and analogously for $q_y$ and $q_z$.  With the product 
\begin{equation}
     \Dtilde_K(\qvec) = D_K(q_x) D_K(q_y) D_K(q_z), 
\end{equation}
this gives the potential of the active particle in the aggregate of the target
particle and its images:
\begin{equation}
U_K = 4 \pi c_\OneP \int_{-\infty}^{\infty}  
\frac{\diff^3 \qvec}{{(2\pi)}^3} \Dtilde _K(\qvec)\,  \expa{i {\qvec} \cdot 
\rvec_{\OneP \TwoP} }\,
\frac{\rho_\TwoP({\qvec})}{|\qvec|^2}.
\label{equ:UatK}
\end{equation}
\Eq{equ:kernel} is the Dirichlet
kernel which converges, in a weak sense, to a sum of $\delta$-functions in the
limit of large $K$:
\begin{equation}
D_K(q_x) \xrightarrow[ K \to \infty]{} \frac{2 \pi}{L} \sum_{m=-\infty}^\infty 
\delta\glb q_x - m \frac{2 \pi}{L}\grb, 
\end{equation}
and similarly for $q_y$ and $q_z$.  The width of the central peak of $D_K$
scales as $1/K$ for large $K$.  Integrals over sufficiently well-behaved
objects become  summations in the limit of large $K$:
\begin{equation}
\int \frac{\diff^3 \qvec}{{(2 \pi)}^3} \tilde D_K(\qvec)   f({\qvec}) 
\rightarrow
\frac{1}{L^3}\sum_{{\qvec} = 2 \pi {\bf m}/L} f({\qvec}) .
\end{equation}
For the volume-charge model, the structure factor is
\begin{equation}
\rho_\TwoP({\qvec}) = c_{\TwoP} \glb 1 - 
\sincf\frac{q_x L}{2} 
\sincf \frac{q_y L}{2}
\sincf \frac{q_z L}{2} 
 \grb, 
\label{equ:rhot}
\end{equation}
where the first term on the right-hand side describes the point charge and the
product of cardinal sine functions, $\sincf(q_x)= \sin(q_x)/q_x$, etc., the
uniform background volume charge.

From \eqtwo{equ:UatK}{equ:rhot}, the  potential of a finite cubic array
of  images of the particle \TwoP, with active particle \OneP, is
\begin{multline}
U_K^{\text{volume}} = 
c_\OneP c_{\TwoP} \int   
\frac{\diff^3  \qvec}{2\pi^2} \tilde D_K(\qvec)   
\frac{
\expa{i  {\qvec}\cdot {\rvec_{\OneP \TwoP}}}   
}{ | \qvec|^2 }
 \\ \times
\glb 1 - \sincf\frac{q_x L}{2} 
\sincf \frac{q_y L}{2} 
\sincf \frac{q_z L}{2} \grb.
\label{eq:Sinc1} 
\end{multline}
For line charges of length $pL$, we find
\begin{equation}
U_K ^{\text{line}}=  c_\OneP c_{\TwoP} \!\! \int  \frac{\diff^3 \qvec}{2\pi^2}  
\Dtilde_K(\qvec)  \frac{\expa{i  {\qvec}\cdot {\rvec_{\OneP \TwoP} }  } }{ 
|\qvec|^2 } \!
       \glb 1 - \sincf \frac{p q_x L}{2} \grb.
\label{eq:Sinc2}
\end{equation}
The volume-charge model is equivalent to the tin-foil Coulomb potential.
The line-charge model, whose potential is not absolutely convergent, is
nevertheless equivalent for ECMC because, as we will see,  the integrals
in \eqtwo{eq:Sinc1}{eq:Sinc2} yield uniquely defined and equivalent Coulomb
derivatives for large $K$.  The difference between the two is given by:
\begin{multline*}
\Delta  U_K = (U_K^{\text{line}}- U_K^{\text{volume}})=  
    c_\OneP c_{\TwoP}  \int\, \frac{\diff^3 \qvec}{2 \pi^2}\, \Dtilde_K(\qvec) 
             \frac{\expa{i {\qvec} \cdot  {\rvec_{\OneP \TwoP}}    }}
             {|\qvec|^2} \\ \times 
           \glb \sincf  \frac{q_x L}{2}  \sincf
  \frac{q_y L}{2}  \sincf  \frac{q_z L}{2}    - 
 \sincf  \frac{ p 
q_x L}{2}  \grb   
\end{multline*}
The Dirichlet kernels imply that the integral in this equation is dominated by
contributions near $\qvec = 2\pi \mvec /L $. However, the function $\sincf(q_i
L/2)$ also has zeros at these same points (except when $ q_i =0$, where the
$\sincf$ function is equal to one).  For large $K$, the potential differences
is thus dominated by a sum over $q_y$, $ q_z$, with $q_x=0$.  This implies
that the potential on the active particle equals (to within a constant)
the tin-foil potential for motion parallel to the line-charges, but the
difference of potentials is corrugated in the perpendicular $y-z$ plane. This
is a consequence of the fusion of multiple aligned line charges into a single
uniform line when $p$ is integer (see \subfig{fig:LineChargesAndGeometry}{b}).

We examine the derivative of $\Delta U_K $ to show that the Coulomb derivatives
converge to the same value:
\begin{multline}
\partpartshort{\Delta U_K}{x_1} = 
\int \, \frac{\diff^3 \qvec}{2 \pi^2}\,   \Dtilde_K(\qvec) 
\frac{q_x \sin(\qvec \cdot  \rvec_{\OneP \TwoP}) }  {|\qvec|^2}  
\\ \times  
          \glb \sincf \frac{q_x L}{2} \sincf
  \frac{q_y L}{2} \sincf \frac{q_z L}{2}    - 
 \sincf  \frac{ p 
q_x L}{2}   \grb   ,
\label{equ:twoDiff} 
\end{multline} 
which suppresses the contributions which remained for the calculation of the
potential, due to the factor  $q_x \sin(q_x x)$ near $q_x=0$.

Finally, we consider explicitly the possible divergence at $ |\qvec | =0$ in
\eq{equ:twoDiff}, due to the presence of the term $1/|\qvec|^2$. We expand all
the trigonometric functions in the integrand, $\Delta I_K$, to find
\begin{equation*}
\Delta I_K \xrightarrow[\qvec \rightarrow 0]{}   \text{const}\times \frac{q_x^2 
\left[
    (p^2-1)q_x^2-q_y^2- q_z^2\right ] }{|\qvec|^2}  \Dtilde_K(\qvec).
\end{equation*}
Even this contribution is thus driven to zero for large $K$. We conclude
that in a periodic three-dimensional system, the line-charge model
becomes equivalent to the volume-charge model, and therefore to tin-foil
electrostatics. The line charges must lie parallel to the direction of motion
but can of course be switched at will. In contrast, the volume-charge model
gives the tin-foil Coulomb derivatives in all directions.

\subsection{Algorithms for Coulomb derivatives}
\label{sec:CoulombDerivatives}
The merged-image Coulomb  derivatives are best  computed from the 
tin-foil
expressions of \eq{equ:metallicEventRate}.  To accelerate the evaluation, we
reduce the Fourier-space component of \eq{equ:DirectDerivFourier} to a sum over
non-negative components $(m_x, m_y, m_z)$:
\begin{equation}
\qtilde_f(\rvec_{\OneP \TwoP}) =  A_{xyz} \sin (\lambda_{12}^x) \cos
  (\lambda_{12}^y) \cos (\lambda_{12}^z) ,
\label{equ:DirectDerivFourierCondensed}
\end{equation}
where $\lambda_{12}^x = 2\pi m_x x_{12} / L$, and similarly in $y$ and $z$ and
where
\begin{equation}
A_{xyz} = \frac{16 c_\OneP c_{\TwoP} m_x}{L^2 | \mvec|^2 2^{\delta_{m_y, 0}
+ \delta_{m_z, 0}}} \exp \left(-\frac{\pi^2 | \mvec|^2 }{\alpha^2 L^2}
\right) 
\end{equation}
is a position-independent tensor that can be computed before the simulation
starts.  In \eq{equ:DirectDerivFourierCondensed}, repeated indices $( x, y, z)$
are summed over non-negative integers $( m_x, m_y, m_z )$.

\begin{figure}[tb]
\includegraphics[width = 
\columnwidth]{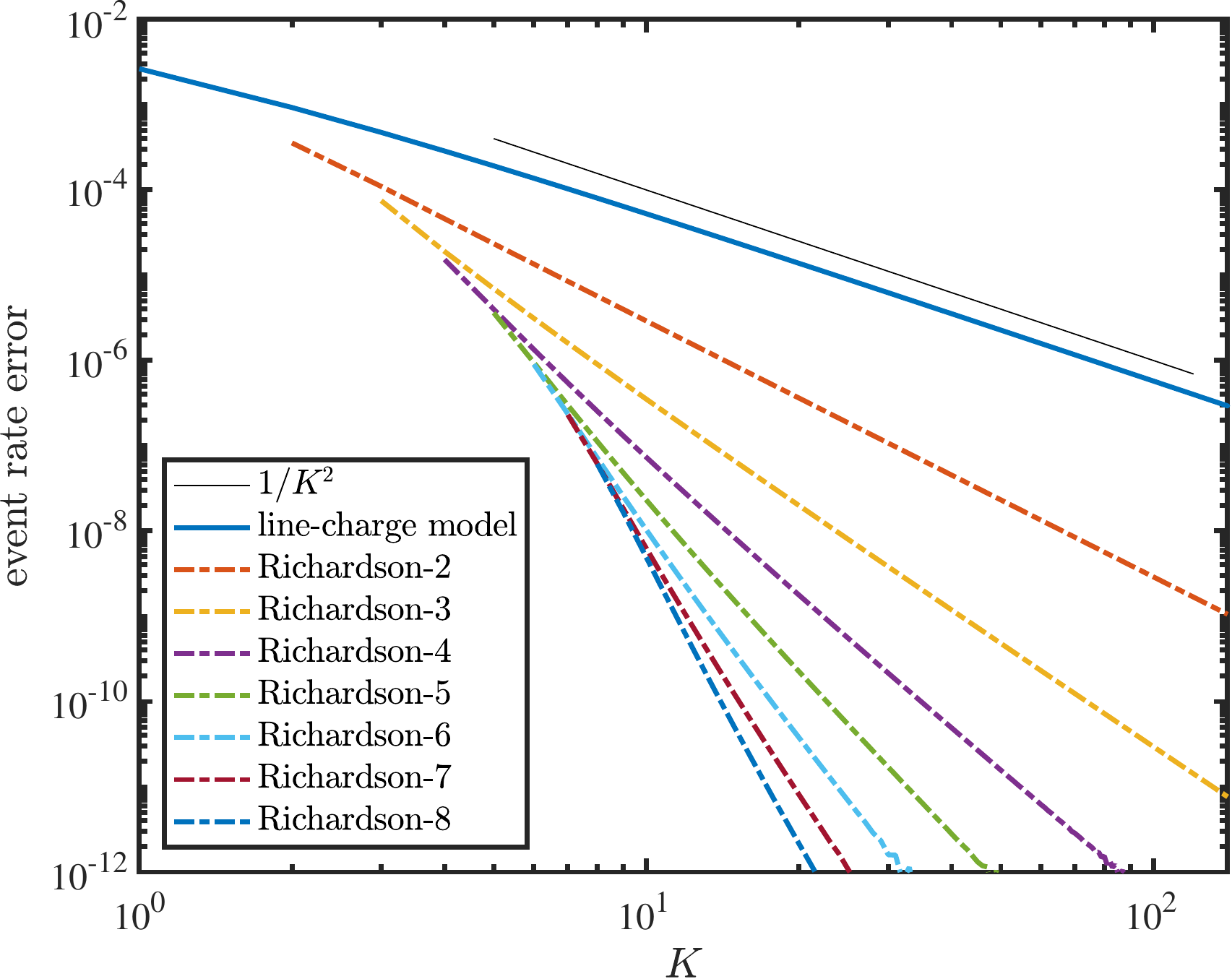}
\caption{Comparison of the tin-foil expression for the Coulomb factor
derivative and the sum over line charges for a given value of $\rvec_{\OneP
\TwoP}$ (see \eq{equ:lineChargeRate}) as a function of the cutoff
$\KK$. The 8-fold iterated Richardson extrapolation for the line-charge 
expression agrees with the tin-foil expression to within $10^{-12}$ for $\KK 
\approx 20$.}
\label{fig:richardson}
\end{figure}

The merged-image Coulomb derivatives can also be computed from the sum 
of the
line-charge derivatives (see the right-hand side of \eq{equ:lineChargeRate}).
Because of the symmetry of the line charges, the quadrupolar contribution
to the derivative is an odd function of $x$, so that forward and backward
terms cancel, and that the sum converges as $1/\KK^2$ for large $\KK$.  The
convergence may be accelerated using Richardson extrapolation~\cite{Bender1978}
(see \fig{fig:richardson}).  Denoting the finite line-charge sum over the range
$ \nvec \in [-\KK,\KK]^3$ as $S_\KK$ and assuming that:
\begin{equation}
S_{\KK} =  S_\infty + \frac{A}{\KK^p}, 
\end{equation}
one may eliminate $A$ as:
\begin{equation}
S'_{\KK +1} = 
\frac{{(\KK+1)}^p  S_{\KK+1}  - {\KK^p} S_{\KK } }
{{(\KK+1)}^p  -  {\KK^p} }.
\label{equ:Richardson}
\end{equation}
The sequence $(S'_{\KK +1} -S_\infty)$ then decays as $1 / \KK^{p+1}$.  The
transformation of \eq{equ:Richardson} can be iterated, each time gaining one
power in the asymptotic behavior of the sequence.  The merged-image line-charge
derivatives converge to the tin-foil expression of \eq{equ:CoulombDirectDeriv},
confirming that the two algorithms compute the same object and that individual
factors in the line-charge model may be used to simulate tin-foil potentials.

As in the line-charge model, one may sum up the associated point charges and
their compensating volume charges explicitly, rather than proceeding through
Fourier transformation. However, the analytic formulas are difficult to work
with. A further possibility consists in compensating each point charge with
more than one line charge. Remarkably, four line charges arranged on a square
of side $L/\sqrt{12}$ in the $y-z$ plane, cancel dipole and quadrupole moments
in the multipole expansion and lead to an absolutely converging sum for the
electrostatic potential.  One may also construct more elaborate sheets and
volumes of screening charges to cancel higher orders in the multipole expansion.
All of these screening objects presented here regularize the sum of the pair
derivatives over images and allow for separate-image factor sets (analogous to
\eq{equ:FactorizedCoulombSec2}, see \sect{sec:CoulombSeparateImage}).  Although
the sequence $S_\KK$ decays faster,  the Coulomb event rate is not reduced by
these different objects.

\subsection{Separate-image ECMC}
\label{sec:CoulombSeparateImage}

As we have seen, all the Coulomb interactions in a finite system with periodic
boundary conditions can be image-merged into a single \COULOMBTYPE\  type
that sums over all the inequivalent minimal paths between two points on a torus, 
and that correspond to images in the rolled-out representation of periodic 
boundary
conditions. For two particles $\OneP$ and $\TwoP$, this is expressed through
a single factor $M = \FACTOR{\SET{\OneP, \TwoP}}{\TYPE{\COULOMBTYPE}}$.
The corresponding factor derivatives can then be computed with the
traditional tin-foil expression (\eq{equ:CoulombDirectDeriv}) or within the
line-charge framework (\eq{equ:lineChargeRate}). The choice of one over the
other is a matter of efficiency only (the algorithmic complexity being the
same). Each of the formulations suggest other choices for the interaction
types. In the line-charge formulation, the choice of an infinite set of
types $\SET{\TYPE[\nvec]{Coulomb}: \nvec \in \ZZ^3}$ suggests itself. For
two particles $\OneP$ and $\TwoP$, the set of separate-image factors is
$\SET{\FACTOR{\SET{\OneP, \TwoP}}{\TYPE[\nvec]{Coulomb}}: \nvec \in \ZZ^3}$.
Within ECMC, these images are statistically independent but only one of
them must be computed precisely for each event. This is because,  as in
\sect{sec:ECMC}, we can use a variant of the cell-veto algorithm (supplemented
with an asymptotic bounding function~\cite{KapferKrauth2016}), in order to
sample the relevant image index $\nvec$ and to then compute the corresponding
factor derivative of $M_\nvec$.

Separate-image Coulomb factors generally come with larger pair event
rates, as the contributions from different images do not compensate (see
\fig{fig:PairEventPlot}). On the other hand, evaluating a separate-image
Coulomb derivative (as in \eq{equ:imagewiseLineChargeDirectDeriv}) to
machine precision requires just a few operations, many fewer than what is
required for its merged-image counterpart.  Details of the separate-image
Coulomb factors can influence the efficiency of the algorithm. As an
example, the terminal point  of the line charge is a singular point of
\eq{equ:imagewiseLineChargeDirectDeriv} and should not approach another point
charge in the system. This motivates our choice of length $2L$ (or multiples
thereof), as the terminal point of one line charge then coincides  with the
position of an image of the original particle. For the Coulomb potential,
the nonphysical line-charge singularity, confounded with the singularity of the
point charge, no longer disrupts the ECMC dynamics.

\begin{figure}[hth]
\includegraphics[width = \columnwidth]
{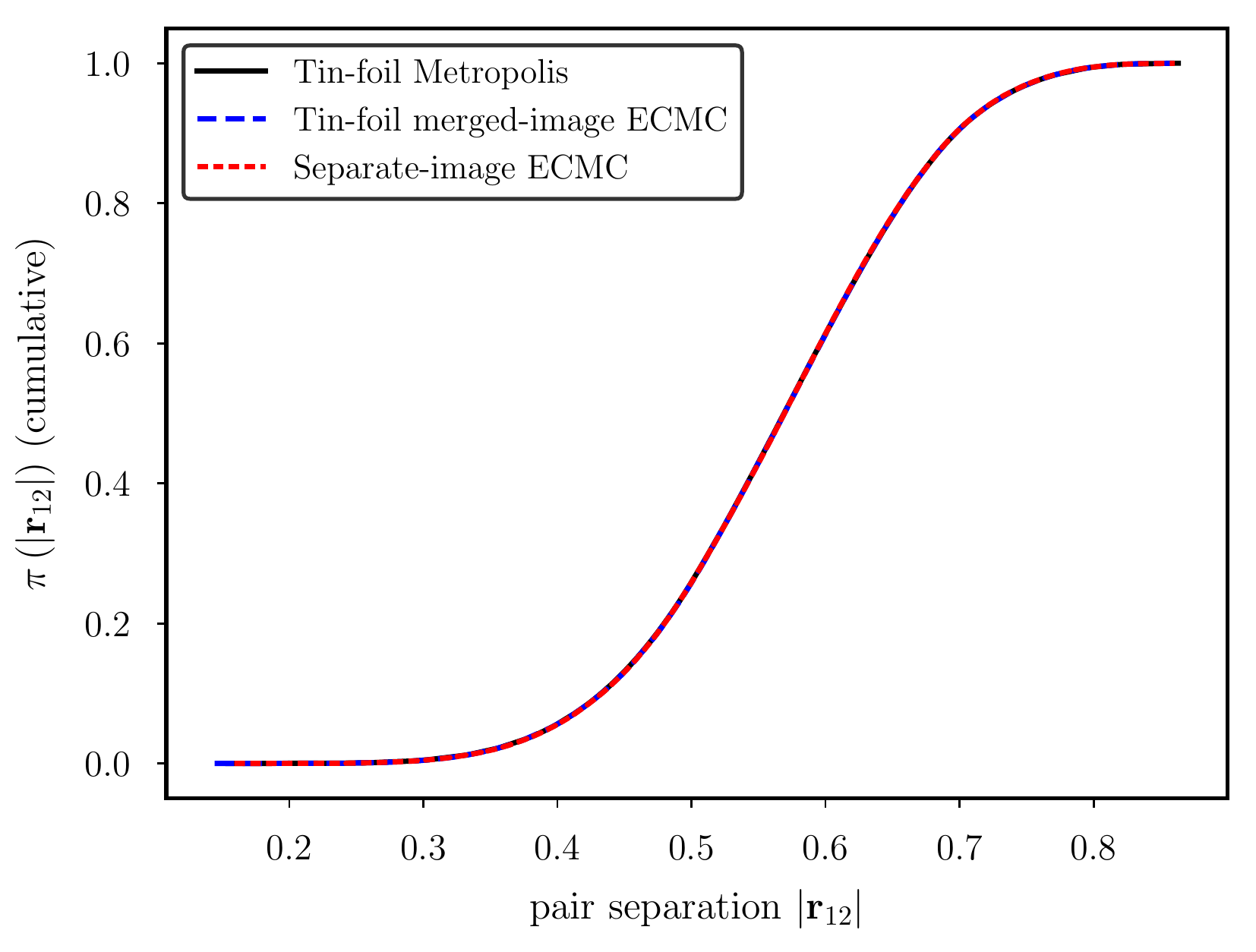}
\caption{Cumulative histogram of the pair separation $ |\rvec_{12}|$ for
two particles of equal charge in a periodic three-dimensional simulation box
($\beta c_1 c_2 = 2$, $L=1$).}
\label{fig:TwoBodyCoulombTorus}
\end{figure}

The dynamic behavior of the different factor sets for the Coulomb problem have
not yet been explored in detail. As a first step, for a system of  two like
Coulomb charges, merged-image and separate-image ECMC was validated against
the regular tin-foil Metropolis algorithm  (see \fig{fig:TwoBodyCoulombTorus}).
All three methods clearly sample the Boltzmann distribution in the asymptotic
steady state.

\section{Dipole--dipole factors}
\label{sec:Dipoles}

\begin{figure}[ht]
\includegraphics{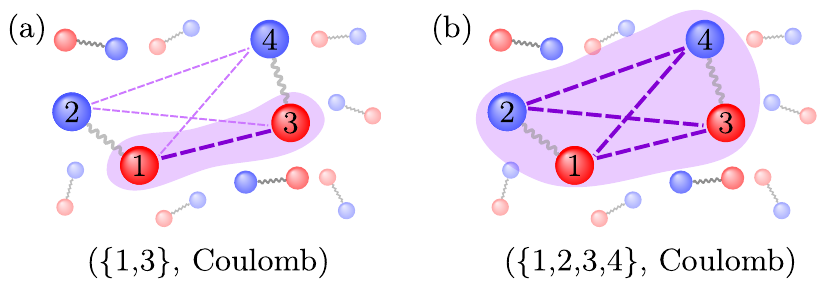}
\caption{Model of two-particle dipoles. \subcap{a}: Particle--particle factor
associating two point charges that belong to different dipoles. \subcap{b}:
\quot{dipole--dipole} factor comprising four Coulomb interactions. }
\label{fig:ManyDipoles}
\end{figure}

In ECMC, one may tailor the factor sets to the problems at hand.  In
electrostatic systems made up of local dipoles, specific \quot{dipole--dipole}
Coulomb factors may thus contain all the atoms distributed over two molecules
that can be far apart from each other.  These factors yield much smaller event
rates than \quot{particle--particle} pair factors. In addition, a special
\quot{inside-first} lifting scheme can direct most of the lifting flow from
the active particle to a target particle situated on the same molecule. Even
for a non-local factor made up of two distant dipoles, the lifting flow will
thus mostly be between an active particle and a target particle on the same
molecule (the probability of an intramolecular lifting grows like $\loga{N}$,
whereas all the intermolecular liftings remain constant). We expect such
a local lifting scheme for extended factors to show interesting dynamic
properties.  In the present section, we explore dipole--dipole factors in a
simple model of charge-neutral two-particle molecules before employing them, in
\sect{sec:Water}, to a model of liquid water. We expect dipole--dipole factors
and their variants to have useful applications in ECMC.

Concretely, for a simple model of two-particle dipoles in a three-dimensional
periodic simulation box, the dipole--dipole factor for the particles
$\SET{\OneP, \TwoP, \ThreeP, \FourP}$ is given by:
\begin{equation}
\FACTOR{\SET{\OneP, \TwoP, \ThreeP, \FourP}}{\TYPE{\COULOMBTYPE} }, 
\label{equ:FactorCoulombDimer}
\end{equation}
(see \subfig{fig:ManyDipoles}{b}), where the corresponding Coulomb 
factor potential is:
\begin{multline}
U_{\FACTOR{\SET{1,2,3,4}} {\TYPE{Coulomb} }}(\rvec_1 \TO \rvec_4) \\
=  \sum_{i = 1}^2
  \sum_{j = 3}^4 U_{\COULOMB}(\rvec_{ij},\SET{c_i, c_j}) .
\label{equ:DipoleCoulombFactorPotentialMolecularFact}
\end{multline}
The factor of \eq{equ:FactorCoulombDimer} thus comprises the
four Coulomb potentials between these particles, using the Coulomb
potential of \eq{equ:CoulombPotentialGeneral}.  The model excludes, as is
usual~\cite{Wu2006}, Coulomb interactions within a dipole.  For the same four
particles, one may also use the \quot{particle--particle} factors
\begin{multline}
\SET{
\FACTOR{\SET{1,3}}{\TYPE{Coulomb}},
\FACTOR{\SET{1,4}}{\TYPE{Coulomb}} \\
\FACTOR{\SET{2,3}}{\TYPE{Coulomb}},  
\FACTOR{\SET{2,4}}{\TYPE{Coulomb}}
 } ,
\label{equ:FactorizationSchemeParticle}
\end{multline}
with the \quot{particle--particle} factor potential:
\begin{equation}
U_{\FACTOR{\SET{i,j}}{\TYPE{Coulomb}}} (\rvec_{ij},\SET{c_i,c_j}) =
  U_{\COULOMB}(\rvec_{ij}, \SET{c_i, c_j})
\label{equ:DipoleCoulombFactorPotentialParticleFact}
\end{equation}
(see \subfig{fig:ManyDipoles}{a}).  We suppose that the particle \OneP\ is
active.  The dipole--dipole event rate
\begin{equation}
\beta
\maxZeroc{\FACDERIV{\FACTOR{\SET{1,2,3,4}}{\TYPE{Coulomb}}}{1} }
\label{equ:DipoleDipoleFactorCoulombEvent}
\end{equation}
then allows the interactions $U_{\COULOMB}(\rvec_{13})$ and
$U_{\COULOMB}(\rvec_{14})$ to compensate each other (and to give the event
rate corresponding to a point charge interacting with a dipole), while the
particle--particle event rate
\begin{equation}
\beta
\maxZeroc{\FACDERIV{\FACTOR{\SET{1,3}}{\TYPE{Coulomb}}}{1}} +
\beta
\maxZeroc{\FACDERIV{\FACTOR{\SET{1,4}}{\TYPE{Coulomb}}}{1}}
\label{equ:DipoleParticleFactorCoulombEvent}
\end{equation}
remains much larger (corresponding to a point charge separately interacting
with two isolated point charges), because the unit-ramp functions are both
non-negative (see \eq{equ:UnitRamp}) and one of them is usually zero.

\subsection{Event-rate scaling for Coulomb factors}
\label{sec:DipoleSystemModel}

\begin{figure*}[ht]
\includegraphics{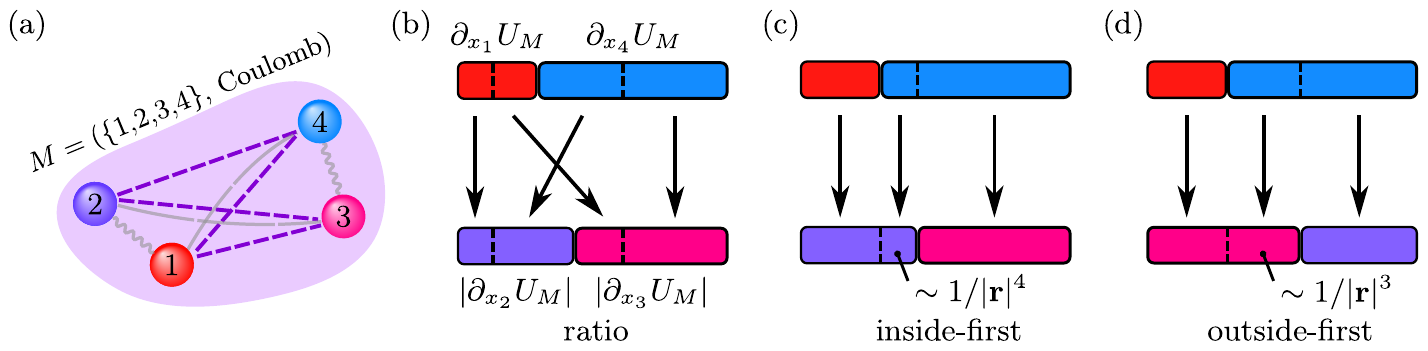}
\caption{Lifting schemes for a dipole--dipole factor. \subcap{a}:
Dipole--dipole factor with four Coulomb interactions. It is assumed
that $\partpartshort{U_M}{x_1}> 0$, $\partpartshort{U_M}{x_4}> 0$, and
$\partpartshort{U_M}{x_2}< 0$, $\partpartshort{U_M}{x_3}< 0$.  \subcap{b}:
\quot{ratio} lifting, \subcap{c}: \quot{inside-first} lifting \subcap{d}:
\quot{outside-first} lifting. }
\label{fig:FactorsDipoles}
\end{figure*}

We now consider a homogeneous system of dipoles of size $|\dvec| \sim d$
small compared to the simulation box (see \fig{fig:FactorsDipoles}). For
concreteness, we suppose that particle $\OneP$ is the active particle. The
event rate, whose scaling with system size we compute in the present
section, is the result of the interaction between the particle $\OneP$ and
the distant dimer (in \fig{fig:ManyDipoles} made up of particles $\ThreeP$
and $\FourP$). As there is no  Coulomb interaction between particles on
the same dipole, the position of particle $\TwoP$ (the dipole partner of
particle $\OneP$) does not come into play for the event rate.  We will see in
\sect{sec:DipoleLiftingSchemes}, that this is no longer true for the lifting
rates, which are influenced both by the distant dimer and by the local dimer of
particle $\OneP$, that is, by the position of particle $\TwoP$.

The electrostatic potential at a distance $\rvec$ from a point charge $c_k$,
within the merged-image (tin-foil) formulation in a box of side $L$, is 
given
by the scaling form:
\begin{equation}
\psi_L(\rvec)  = \frac{c_{k} }{|\rvec|}  f_E(\rvec/L), 
\label{equ:scalingForm}
\end{equation}
which generalizes Coulomb's law valid in free space.  The function
$f_E(\xvec)$, is smooth and remains $\bigO{1}$ for all $\xvec \in
{[-1/2,1/2]}^3$.  For separations such that $|\rvec|/L\ll 1$ the potential
given by \eq{equ:scalingForm} has the expansion~\cite{fraserEwaldExpansion}:
\begin{equation}
\psi(\rvec) = c_{k} \glb \frac{1}{|\rvec|} +\frac{\text{const}}{L} + 
\frac{2 
\pi
    |\rvec|^2}{3L^3}
+\dots \grb .
\label{equ:expandEwald}
\end{equation}
The
$n$th-order derivatives of $f_E(\rvec /L)$ are also smooth and have an
amplitude which scale as $L^{-n}$.  The Coulomb derivative between an active
particle and a particle $k$, separated by a vector $\rvec \in {[-L/2,L/2]}^3$,
also has the scaling form:
\begin{equation}
\beta \FACDERIV{\FACTOR{\SET{1,k}}{\TYPE{\COULOMBTYPE} } }{1}  =   
\frac{\BJERR  }{|\rvec| ^2}  f^{1}_E(\rvec / L). 
\label{equ:ScaleOne}
\end{equation}
Here, we have introduced the characteristic Bjerrum length $\BJERR =
|e^2| \beta$, with $e$ the elementary charge, the distance at which the
Coulomb interaction equals the thermal energy and used $f_E^1$ as a new
scaling function, which again remains $\bigO{1}$.  An explicit form for
\eq{equ:ScaleOne} at small separations can be found from \eq{equ:expandEwald}.
For a constant number density $\rho$ of particles within the simulation cell,
the mean total Coulomb event rate per particle, $\mean{Q_{\text{p--p}}}$, is
given by the integral:
\begin{align}
\mean{Q_{\text{p--p}}}= & \sum_{k \ne 1} 
\mean{q_{\FACTOR{\SET{1,k}}{\TYPE{\COULOMBTYPE}}}                         
    } \\
= & \int_{{[-L/2,L/2]}^3} \frac{\BJERR \rho}{|\rvec|^2} 
f^{1}_E(\rvec/L) \, \diff^3
  \rvec  \nonumber \\  =&  \BJERR \rho L\int_{{[-1/2,1/2]}^3} 
\frac{1}{\xvec^2}
                  f^{1}_E(\xvec) \, \diff^3 \xvec \sim {\BJERR \rho L} .
\label{equ:Qpp}
\end{align}
This mean total event rate thus diverges as \bigO{L}. The inverse of
$\mean{Q_{\text{p--p}}}$ sets the scale for the mean-free path due to
charge--charge interactions, and it is of length scale \bigO{1/L}. The result
agrees with the naive free-space argument~\cite{KapferKrauth2016} based on the
bare $1/|\rvec|$ Coulomb interaction.  At constant density, the divergence of
\eq{equ:Qpp} in $L\sim N^{1/3}$ implies that the active and target particles
are often widely separated from each other.  With pair factors, one thus
expects a complexity of \bigO{N^{4/3}} for an \bigO{1} displacement of all
particles in the system.

The scaling form of the potential can also be used to determine the event rate
for dipole--dipole factors (as in \subfig{fig:ManyDipoles}{b}), the interaction
of point charges with dipoles, or the interaction of pairs of well-separated
dipoles.  The potential at a distance $\rvec$ from  a dipole in the periodic
box is found from \eq{equ:scalingForm} by applying the operator $(-\dvec\cdot
\nabla)$, with $\dvec$ the dipole moment.  Using again $|\dvec| \sim d$,
this implies that the event rate of the dipole--dipole factor, resulting
from the interaction of the active particle $\OneP$ with the dipole at  a
distance $\rvec$ corresponds to a particle--dipole Coulomb interaction. The
dipole--dipole event rate, for two dipoles separated by a vector $\rvec \in
{[-L/2,L/2]}^3$ is given by:
\begin{equation}
\beta \FACDERIV{\FACTOR{\SET{1,2,3,4}}{\TYPE{\COULOMBTYPE} } }{1}
  \sim \frac{d \BJERR}{|\rvec|^3} f^1_E (\rvec/L), 
\label{equ:DipoleEventRateScaling1}
\end{equation}
where $\rvec$ denotes the vector from the active particle to the dipole. 
\Eq{equ:DipoleEventRateScaling1} implies that ECMC with dipole--dipole factors 
has a much lower mean total Coulomb event rate
$\mean{Q_{\text{Coulomb}}}$:
\begin{multline}
\mean{Q_{\text{Coulomb}}} =   \int_{{[-L/2,L/2]}^3} \frac{\BJERR \rho 
d }{|\rvec|^3} 
         f^{2}_E(\rvec/L) \, \diff^3
  \rvec   \\ = \BJERR \rho d \int_{{[-1/2,1/2]}^3} \frac{1 
}{|\xvec|^3}
                           f^{2}_E(\xvec) \, \diff^3 \xvec ,
\label{equ:LogN}
\end{multline}
where $  f^{2}_E$ is another scaling function.
The second integral in \eq{equ:LogN} is weakly divergent near the origin
(which simply means that in ECMC very nearby dipoles have to be treated
individually). Excluding a region of radius $\bigO{d / L}$, the mean  total
Coulomb event rate using dipole--dipole factors is
\begin{equation}
\mean{Q_{\text{Coulomb}} } \sim \BJERR \rho d \logb{L/d} .
\label{equ:LogN2}
\end{equation}
This much reduced total event rate, obtained by limiting the contributions from 
large distances, is our main motivation for using dipole--dipole factors. 

The scaling obtained in \eqtwo{equ:DipoleEventRateScaling1}{equ:LogN2} is
independent of the specific definition of the dipole model. It only relies on
the use of dipole--dipole factors connecting two charge-neutral molecules that
may be far apart (see \sect{sec:Water}, where the dipoles are realized by \hoh\
molecules). The scaling is also insensitive to the introduction of screening
charge distributions, and it holds both for the merged-image and for the
separate-image factor sets.  Adapting this factorization framework to systems
composed of molecules that behave as approximate higher-order multipoles would
further improve the scaling.

\subsection{Dipole--dipole lifting schemes}
\label{sec:DipoleLiftingSchemes}

We now consider lifting schemes for dipole--dipole factors, and for
concreteness, we consider a four-particle system of particles $\SET{\OneP,
\TwoP}$, forming a charge-neutral dipole $\dvec_{12}$ and particles
$\SET{\ThreeP, \FourP}$, forming an analogous dipole $\dvec_{34}$.  In this
two-dipole system, particle \OneP, for example,  not only interacts with a
charge-neutral dipole $\dvec_{34}$, but is itself inside such a dipole 
$\dvec_{12}$. Although the Coulomb lifting rate is oblivious to
the position of $\TwoP$ (as there is no Coulomb interaction between particles
$\OneP$ and $\TwoP$), particle \TwoP\ is part of the dipole--dipole factor, and
its position influences the relative lifting rates.

We obtain the  derivatives with respect to particles \OneP\ and \TwoP\ for the
factor $M = \FACTOR{\SET{1,2,3,4}}{\TYPE{\COULOMBTYPE}}$ as follows:
\begin{align}
\beta \FACDERIV{M}{1}=& 
            \BJERR \glc \dip \frac{ |\dvec_{34}|}{ |\rvec|^3} + 
              \bigOb{\frac{|\dvec_{34}|^2}{|\rvec|^4} } 
+\bigOb{\frac{|\dvec_{34}|}{L^3} } \grc,\label{equ:q1}\\
\beta \FACDERIV{M}{2}=& 
           \BJERR\glc -\dip \frac{|\dvec_{34}| }{|\rvec|^3} + 
\bigOb{\frac{|\dvec_{34}|^2}{|\rvec|^4}}+\bigOb{\!\frac{|\dvec_{34}|}{L^3}\!} 
\grc. \label{equ:q2} 
\\
\intertext{The dominant terms in these two equations are equal in magnitude
yet opposite in sign, reflecting that particles $\OneP$ and $\TwoP$ interact
with the same distant dipole $\dvec_{34}$, are of opposite sign, and close to
each other (on the dipole $\dvec_{12}$).  For the factor derivatives with
respect to particles $\ThreeP$ and
$\FourP$, we find:}
\beta \FACDERIV{M}{3}=& 
\BJERR \glc \diptilde \frac{|\dvec_{12}|}{|\rvec|^3} + 
\bigOb{\frac{|\dvec_{12}|^2}{|\rvec|^4}}+\bigOb{\frac{|\dvec_{12}|}{L^3}} \grc,
\label{equ:q3} 
\\
\beta \FACDERIV{M}{4}=& 
\BJERR \glc -\diptilde \frac{|\dvec_{12}|}{ |\rvec|^3} + 
  \bigOb{\frac{|\dvec_{12}|^2}{|\rvec|^4}}+
  \bigOb{\! \frac{|\dvec_{12}|}{L^3} \!} \grc. 
\label{equ:q4}
\end{align}
(For ease of notation, we used here \eq{equ:expandEwald} for small $|\rvec|/L$
rather than the full scaling form.)

The coefficient $\dip$ (and analogously for $\diptilde$) reflects the
orientation of $\dvec_{\ThreeP \FourP}$ with respect to the distance vector
between the two dipoles (see \fig{fig:FactorsDipoles}).  Remarkably,
the factor derivatives of $M$ with respect
to the particles  within each dipole ($\FACDERIV{M}{1} +\FACDERIV{M}{2}$
and $\FACDERIV{M}{2} + \FACDERIV{M}{4}$) cancel at order $1/|\rvec|^3$ and
leave a remainder of $1 / |\rvec| ^4$.  This dipole--dipole compensation to
order $1/|\rvec|^3$ of the factor derivatives is a general feature for
pairs of local dipoles (that can be composed of more than two atoms) inside a
factor, and occurs in the same manner with the full scaling functions in the
merged-image potential.

\begin{table}[htb]
\begin{tabular}{l|c|c|c}
    Lifting scheme & $q_{\text{intra}}\,\, 
                      q_{\text{inter}} $ &
                     $ \mean{Q_{\text{intra}}}\,\,\, 
 \mean{Q_{\text{inter}}}$  & Lifting \\ \hline
    particle & $ \,0$\, \, \, \, $1/|\rvec|^2$ & 0\,\, $L $&  inter-dipole \\ 
\hline
    ratio & $ 1/ |\rvec|^3$\, $1/|\rvec|^3$ & $\log L\,\, \log L$& inter+intra 
\\
    outside-first & $1 / |\rvec|^3 $\, $1 / |\rvec|^3$ & $\loga{L}\,\, 
\loga{L}$& inter+intra \\
    inside-first & $ 1 / |\rvec|^3$\, $1 / |\rvec|^4$ & $\loga{L} \,\,\, \const 
$& intra-dipole\\\hline
\end{tabular}
\caption{Coulomb lifting rates for two dipoles separated by a distance
$|\rvec|/L \ll 1 $, together with full integrated rate in simulation
box of size $L^3$: One particle--particle and three dipole--dipole
schemes (\quot{ratio}, \quot{outside-first} and \quot{inside-first}).
$q_{\text{intra}}$: lifting rate to the non-active particle within the active
dipole. $q_{\text{inter}}$: lifting rate to the triggering dipole.  $\mean{
Q_{\text{intra} }}$ and $\mean{ Q_{\text{inter}}}$ denote the mean total
event rates (using the full scaling form, as in \sect{sec:DipoleSystemModel}),
integrated over the simulation box.}
\label{tab:TabFourBody}
\end{table}

We recall from \eq{equ:DerivativesBalance} that the four factor derivatives
exactly sum up to zero. As illustrated in \sect{sec:ECMCLifting} (see
\fig{fig:FactorBasics}), the lifting scheme corresponds to arranging the
indices $k^+ \in I_M^+$ on the upper row of a two-row table and the  indices
$k^- \in I_M^-$ on the lower row. In a factor with large separation $|\rvec|$,
each row contains one element corresponding to each of the two dipoles (see
\fig{fig:FactorsDipoles}).

The \quot{ratio} lifting scheme is as described in \sect{sec:ECMCLifting}.  All
elements fall off as $\bigO{1/|\rvec|^3}$ (see 
\eq{equ:DipoleEventRateScaling1}), and both rows
contain elements representing each dipole. From \eqtwo{equ:q2}{equ:q4}, this
leads to comparable proportions of intra- and inter-molecular liftings. Both
rates fall off at the same rate, but their coefficients are different
reflecting the orientations of the dipoles.  The total inter- and intra-dipole
lifting rates both scale as $\log{L}$ (see \subfig{fig:FactorsDipoles}{b} and
\tab{tab:TabFourBody}).

In the \quot{inside-first} lifting scheme, the elements corresponding to each
dipole are aligned with each other. The two match to order $\sim 1/ |\rvec|^3$.
The mismatch in bar length is $\bigO{1/|\rvec|^4}$ in \eqtwo{equ:q1}{equ:q2}.
In the full scaling picture, the difference in length of the elements can be
computed analogously. Coulomb liftings thus occur mostly within a dipole, and
long-ranged inter-dipole liftings remain bounded in number for large system
sizes (see \subfig{fig:FactorsDipoles}{c} and \tab{tab:TabFourBody}).

Finally, the \quot{outside-first} lifting scheme consists in vertically
aligning elements corresponding to different dipoles. Aligned elements are of
length $\sim |\dip|$ and $\sim |\diptilde|$, so that intra- and inter-dipole
lifting rates again both fall off as $\bigO{1/|\rvec|^3}$. The situation is
analogous to the one for the \quot{ratio} lifting, and the \quot{outside-first}
scheme remains strongly non-local (see \subfig{fig:FactorsDipoles}{d} and
\tab{tab:TabFourBody}).

In contrast to the above dipole--dipole factors, the \quot{particle--particle}
factor, as argued in \eqtwo{equ:ScaleOne}{equ:Qpp}, produces events which
occur at the scale of the simulation box at a rate which decreases as only
$1/|\rvec|^2$, leading to a total event rate increasing linearly with $L$. The
lifting flow is between one  dipole and the other, and the intra-dipole lifting
rate is zero (see \tab{tab:TabFourBody}).

\subsection{Validation of factors and liftings}
\label{sec:DipoleNumericalTests}

The dipole--dipole factors and their different lifting schemes can be checked
for consistency for two charge-neutral dipoles with a short-ranged vibrational
intra-dipole potential, a repulsive potential between oppositely charged
particles (needed to keep dipoles apart from each other) as well as 
intermolecular Coulomb interactions. With particles numbered as in 
\fig{fig:ManyDipoles}, the model corresponds to a factor set
\begin{multline}
\SET{
\FACTOR{\SET{1,2}}{\TYPE{bond}}, 
\FACTOR{\SET{3,4}}{\TYPE{bond}},
\FACTOR{\SET{1,4}}{\TYPE{rep}},\\
\FACTOR{\SET{2,3}}{\TYPE{rep}},
\FACTOR{\SET{1,2,3,4}}{\TYPE{Coulomb}}
 } ,
\label{equ:FactorizationSchemeDipoleIntra}
\end{multline}
with the harmonic bond factor potential, 
\begin{equation}
U_{\FACTOR{\SET{i,j}}{\TYPE{\BONDTYPE}}}(\rvec_{ij}) = \frac12 k_b \glb
  |\rvec_{ij}| - r_0
\grb^2 ,
\label{equ:FactorPotenialDipoleHarm}
\end{equation}
with $k_b > 0$, a short-range repulsive potential
\begin{equation}
U_{\FACTOR{\SET{i,j}}{\TYPE{rep}}}(\rvec_{ij}) 
      =  \frac{1}{2} k_2 \glb \frac{r_0}{|\rvec_{ij}|} \grb ^6, 
\label{equ:FactorPotenialDipoleRep}
\end{equation}
with $k_2> 0$, and a scalar separation
$r_0$, in addition to the 
dipole--dipole Coulomb factor potential of 
\eq{equ:DipoleCoulombFactorPotentialMolecularFact}.

The dipole--dipole Coulomb factor differs from the particle--particle
Coulomb factors in the set:
\begin{multline}
\SET{
\FACTOR{\SET{1,2}}{\TYPE{bond}}, 
\FACTOR{\SET{3,4}}{\TYPE{bond}},
\FACTOR{\SET{1,4}}{\TYPE{rep}},\\
\FACTOR{\SET{2,3}}{\TYPE{rep}},
\FACTOR{\SET{1,3}}{\TYPE{Coulomb}},
\FACTOR{\SET{1,4}}{\TYPE{Coulomb}} \\
\FACTOR{\SET{2,3}}{\TYPE{Coulomb}},  
\FACTOR{\SET{2,4}}{\TYPE{Coulomb}}
} ,
\label{equ:FactorizationSchemeDipoleIntra2}
\end{multline}
where the factor potentials corresponding to bond vibrations and the repulsion 
between unlike charges are as in 
\eqtwo{equ:FactorPotenialDipoleHarm}{equ:FactorPotenialDipoleRep} and the
Coulomb factor potentials are those of
\eq{equ:DipoleCoulombFactorPotentialParticleFact}.
In addition, since $|I_M| = 2$ for each
particle-factorized factor $M$, we have no freedom in choosing a lifting scheme
(see \sect{sec:DipoleLiftingSchemes}).

\begin{figure}[htb]
\includegraphics[width = \columnwidth]{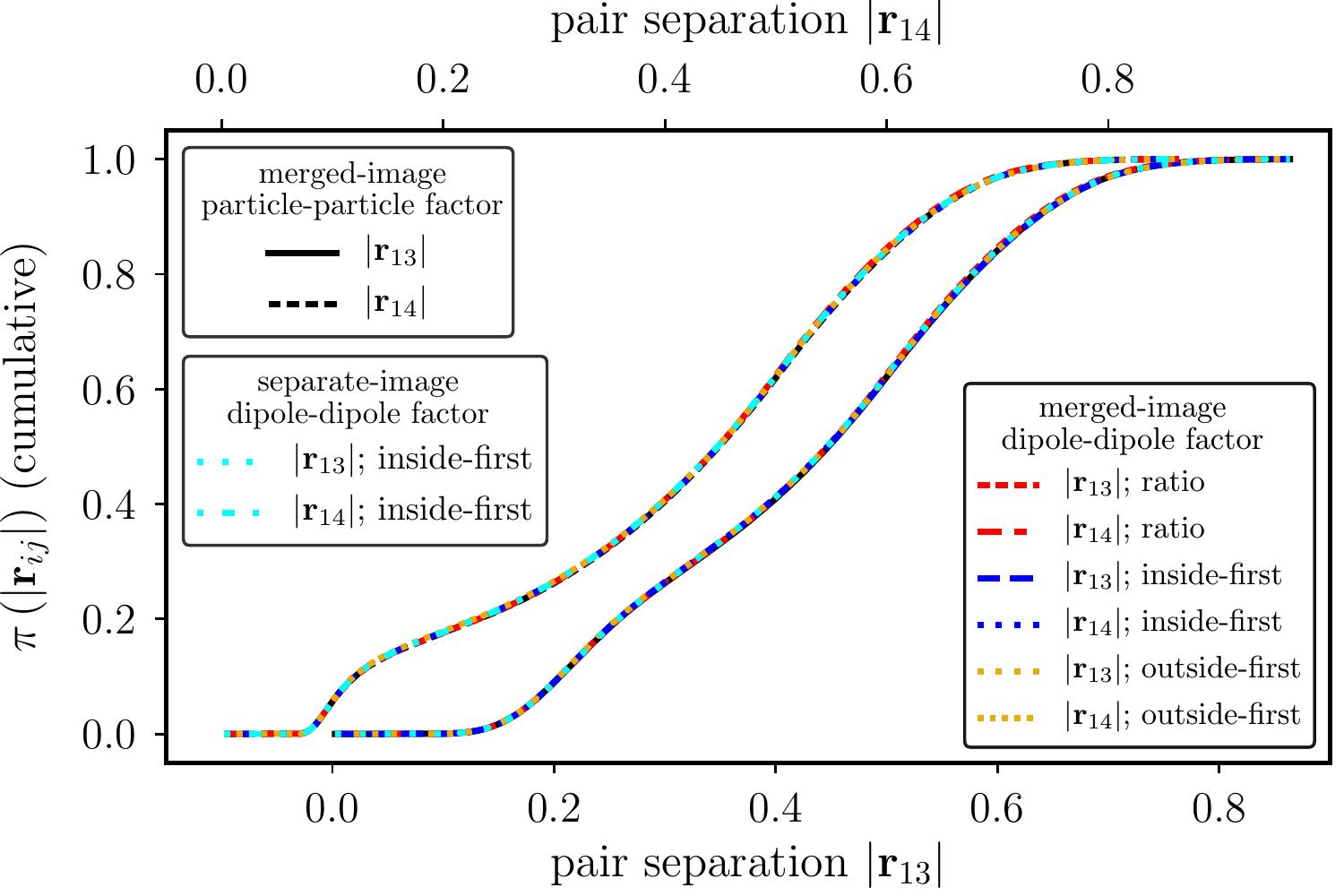}
\caption{Cumulative histograms of the distances $ |\rvec_{13}|$ (like charges,
see \subfig{fig:FactorsDipoles}{a}) and
 $ |\rvec_{14}|$ (opposite charges, see \subfig{fig:FactorsDipoles}{a})
for the particle--particle factor set of
\eq{equ:FactorizationSchemeDipoleIntra2}, and also  for the factor set
of \eq{equ:FactorizationSchemeDipoleIntra} using dipole--dipole Coulomb
factors, using the three lifting schemes of \subfig{fig:FactorsDipoles}{b-d}.
Also separate-image dipole--dipole factors with inside-first lifting.  Periodic
cubic simulation box with $L=1$, $c_i = \pm 1$ point charges, $\beta = 1$, $k_b
= 400$, $k_2 = 1$ and $r_0 = 0.1$. }
\label{fig:DipoleCDF}
\end{figure}

The \quot{ratio}, \quot{inside-first} and \quot{outside-first} lifting
schemes for the dipole--dipole factor are easily implemented and compared to
the particle--particle lifting scheme. By construction, they yield identical
thermodynamic correlations (see  \fig{fig:DipoleCDF}).  Although the event
rates are fixed by the decomposition of the total potential into factors, the
different lifting schemes may differ in their dynamical behavior.

\section{Liquid water and dipole--dipole factors}
\label{sec:Water}

\begin{figure*}[t]
\includegraphics{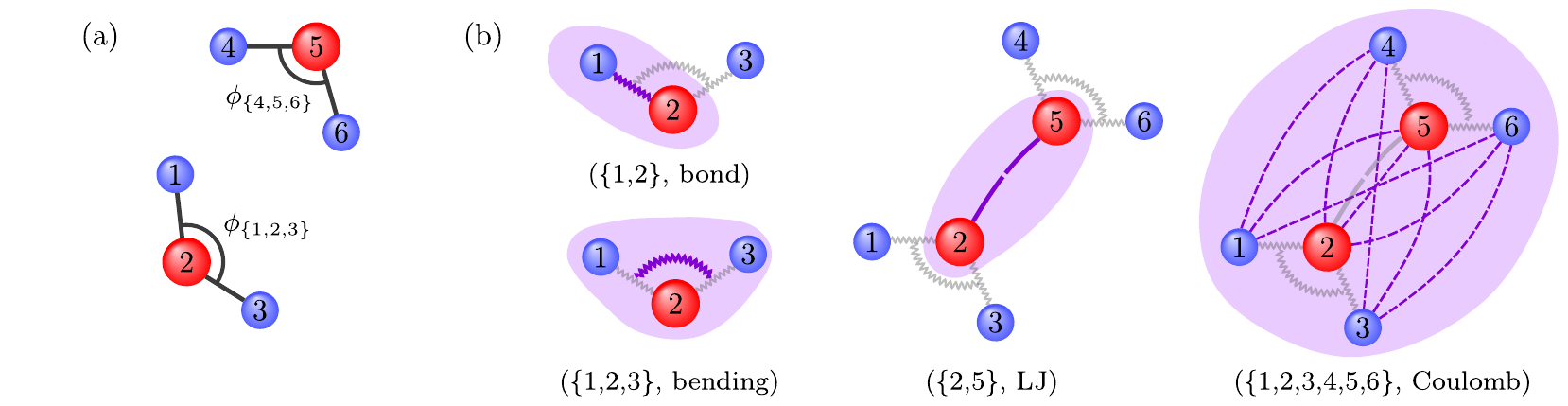}
\caption{SPC/Fw water model and ECMC factors.  \subcap{a}: Two \hoh\
molecules, with particles $\SET{1,2,3}$ and $\SET{4,5,6}$, respectively
($2$ and $5$ being the oxygens). Each of the molecules has a finite dipole
moment.  \subcap{b}: \quot{bond}, \quot{bending}, \quot{LJ} and Coulomb
factors implementing the  SPC/Fw model.  Factors contain between two and six
particles.}
\label{fig:MolFactors}
\end{figure*}

To explore ECMC in a realistic context, we implement in this section the 
SPC/Fw liquid-water model~\cite{Wu2006}. This model
combines the long-ranged Coulomb potential with hydrogen--oxygen bond-length
vibrations, a flexible hydrogen--oxygen--hydrogen angle, and a specific 
oxygen--oxygen interaction
of the Lennard-Jones type. The SPC/Fw model is closely related to one used in
molecular-dynamics simulations of solvated peptides~\cite{Shaw2010}.

Naturally, each water molecule is charge-neutral and dipolar, so that the
dipole--dipole factorization of \sect{sec:Dipoles} applies.  This realizes a
mean free path for a single particle as $\sim 1/\loga{N}$ in the thermodynamic
limit.  (An earlier ECMC Coulomb algorithm~\cite{KapferKrauth2016} had obtained
a mean-free path of as $\sim 1/N^{1/3}$.)

\subsection{Factors in the SPC/Fw water model}
\label{sec:WaterTIP}

To simulate liquid water with the SPC/Fw potential, we use the following type
set:
\begin{equation}
 \SET{\TYPE{\BONDTYPE}, \TYPE{\BENDINGTYPE}, \TYPE{\LJTYPE}, 
\TYPE{\COULOMBTYPE}}.
\label{equ:TypeTIP3P}
\end{equation}
As an example, the factor set for two water molecules, containing particles
$\SET{1,2,3}$ and $\SET{4,5,6}$, respectively, (and with $2$ and $5$ being the
oxygens, see \subfig{fig:MolFactors}{a}) is:
\begin{multline}
\SET{\FACTOR{\SET{1,2}}{\TYPE{\BONDTYPE}}, 
\FACTOR{\SET{2,3}}{\TYPE{\BONDTYPE}}, \\
 \FACTOR{\SET{4,5}}{\TYPE{\BONDTYPE}}, \FACTOR{\SET{5,6}}{\TYPE{\BONDTYPE}}, 
 \FACTOR{\SET{2,5}}{\TYPE{\LJTYPE}},   \\
 \FACTOR{\SET{1,2, 3}}{\TYPE{\BENDINGTYPE}}, 
\FACTOR{\SET{4,5,6}}{\TYPE{\BENDINGTYPE}}, \\
 \FACTOR{\SET{1\TO 6}}{\TYPE{\COULOMBTYPE}} 
}.
\label{equ:FactorizationSchemeWater}
\end{multline}
This factor set (see \subfig{fig:MolFactors}{b}) trivially generalizes to more
than two \hoh\ molecules.

In \eq{equ:FactorizationSchemeWater}, the \quot{bond} factor potential
of \eq{equ:FactorPotenialDipoleHarm} describes oxygen--hydrogen bond vibrations
with the equilibrium bond distance $r_0 =\SI{1.012}{\angstrom}$ and
$k_b=1059.162\,\mathrm{kcal}\, \mathrm{mol}^{-1}\,\mathrm{rad}^{-2}$, 
that correspond to the SPC/Fw parameters. The
\quot{\BENDINGTYPE} factor potential describes the fluctuations in the bond
angle within each \hoh\ molecule:
\begin{equation*}
U_{\FACTOR{\SET{i,j,k}}{\TYPE{\BENDINGTYPE}}}(\rvec_i, \rvec_j, \rvec_k) =
  \frac{1}{2}k_a \glb \phi_{\SET{i,j,k}} - \phi_0 \grb^2 ,
\label{equ:FactorPotenialWaterAngle}
\end{equation*}
where $\phi_{\SET{1,2,3}}$ and $\phi_{\SET{4,5,6}}$ denote the internal angle
between the two legs of each \hoh\ molecule (see \fig{fig:MolFactors}).
We adopt the  SPC/Fw parameters: $\phi_0=113.24^{\circ}$ and
$k_a=75.90\,\mathrm{kcal}\,\mathrm{mol}^{-1}\,{\text \AA}^{-2}$.  The specific
Lennard-Jones interaction between oxygen atoms corresponds to the \quot{LJ}
factor potential
\begin{equation}
U_{\FACTOR{\SET{2,5}}{\TYPE{\LJTYPE}}}(\rvec_{25}) = k_{\text{\LJTYPE}}\glc
\glb\frac{\sigma}{|\rvec_{25}|} \grb^{12} -
\glb\frac{\sigma}{|\rvec_{25}|}\grb^{6} \grc,
\end{equation}
where $k_{\text{\LJTYPE}} = 0.62\,\text{kcal}\,\text{mol}^{-1}$ and $\sigma
= 3.165\,\Ags$ are prescribed in the SPC/Fw model. The Lennard-Jones
interaction is truncated beyond $9.0\,\Ags$.  Finally, the dipole--dipole
\quot{\COULOMBTYPE} factor potential, in direct generalization of
\eq{equ:DipoleCoulombFactorPotentialMolecularFact}, is given by:
\begin{multline}
U_{\FACTOR{\SET{1 \TO 6}} {\TYPE{\COULOMBTYPE} }}  
(\rvec_1 \TO \rvec_6)  \\ =  \sum_{i = 1}^3
  \sum_{j = 4}^6 U_{\COULOMB}(\rvec_{ij},\SET{c_i, c_j}).
\label{equ:DipoleCoulombFactorPotentialMolecularFact3}
\end{multline}
Here, the Coulomb potential of \eq{equ:CoulombPotentialGeneral} is used with
the SPC/Fw parameters $ c_1= c_3=  c_4= c_6= 0.41e$  and  $c_2= c_5 = -0.82e$
(with $e$ the elementary charge).

The type set of \eq{equ:TypeTIP3P} is by no means unique. We could also
break up the Lennard-Jones interaction into two types, corresponding
to the two components of the Lennard-Jones potential (as discussed in
\sect{sec:ECMCFactorized}). Also, instead of the merged-image Coulomb type
we could adopt any of the variants of the separate-image type, resulting in a
type set:
\begin{equation}
\SET{\TYPE{\BONDTYPE}, \TYPE{\BENDINGTYPE}, \TYPE{\LJTYPE},
\TYPE[\nvec]{\COULOMBTYPE} : \nvec \in \ZZ^3} \notag .
\end{equation}
Finally, it is possible to break up the \quot{\BONDTYPE} and
\quot{\BENDINGTYPE} factors into $N_{\text{\hoh}}-1$ equal terms in order to
construct a unique dipole--dipole factor for each pair of \hoh\ molecules  in
such a way that the type set only contains a single element. All these choices
are correct, but they may differ in the ease of implementation and in the speed
with which they approach equilibrium.

\subsection{Intrinsic rotations}
\label{sec:WaterIntrinsicRotations}

Our version of ECMC is formulated in terms of displacements that, for a given
event chain, are along one of the directions $\SET{\UnitX, \UnitY, \UnitZ}$.
Each individual event chain can strain the system, but is unable to rotate
it, as the coordinates perpendicular to the direction of motion remain
unchanged. The flexible SPC/Fw \hoh\ molecule may itself get strained in a
single event chain.  Applying strain subsequently in different directions is
known to be equivalent to a rotation on all levels, and in particular on the
level of a single molecule. This guarantees that the algorithm is irreducible,
and can attain all of configuration space.

The rotation that is induced through subsequent event chains in the three
directions can be illustrated in an ECMC simulation of a single \hoh\ molecule,
using only the intramolecular factor types in \eq{equ:TypeTIP3P}. The
rotational dynamics of such a single molecule is easily tracked through the
equilibrium autocorrelation function of the dipole moment $\dvec = \rvec_{\TwoP
\OneP}+\rvec_{\TwoP \ThreeP}$ (see \fig{fig:MolFactors}), given by
\begin{equation*}
A(s)= \mean{ \dvec(s')\cdot\dvec(s'+s)} \sim \expb{- s / \lambda}\quad 
\text{for $s \to \infty$}, 
\end{equation*}
where the variables $s$ and $s + s'$ denote the ECMC displacement (proportional
to the time of the continuous Markov process).  $A(s)$ decays exponentially
at large $s$ with a rate that gives the autocorrelation length $\lambda$ of
molecular orientation.

\begin{figure}[htb]
\includegraphics[width=\columnwidth]{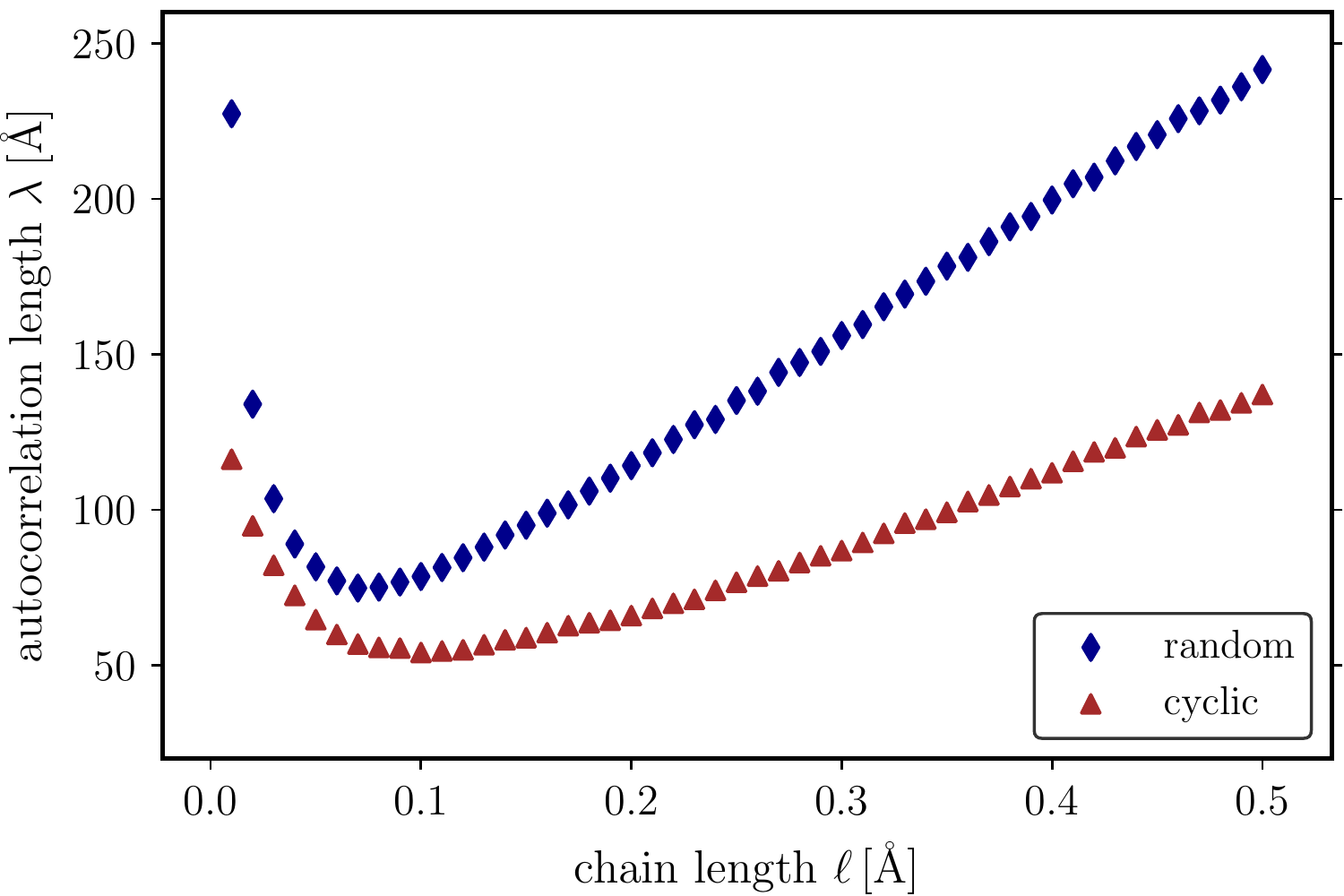}
\caption{Autocorrelation length $\lambda$ for the dipole moment in ECMC  of a
single \hoh\ molecule (fixed chain length $\ell$) for the cyclic sequence of
event-chain directions ($\UnitX, \UnitY, \UnitZ, \UnitX,\dots$) and for their
random resampling.}
\label{fig:RotationsMolecule}
\end{figure}

At temperature $\SI{300}{K}$, the cumulative chain length it takes to rotate
the molecule around itself is about one to two orders of magnitude larger than
the \hoh\ molecule itself (see \fig{fig:RotationsMolecule}). In the limit of
large chain lengths $\ell$, the autocorrelation length of the dipole moment
is proportional to $\ell$. This simply means that lengthening an already long
chain does not add to the internal strain of the water molecule, as a local
equilibrium is reached.

The sequence of chain directions need not be random: The switching of
directions merely renders the Markov chain irreducible, whereas global
balance is satisfied for any infinitesimal move (without the return move
necessary for detailed balance). As a deterministic sequence $\UnitX \UnitY
\UnitZ \UnitX \UnitY\dots$ avoids repetitions, we find it to decorrelate the
dipole moment faster than a uniform random sampling of chain directions (see
\fig{fig:RotationsMolecule}).  The rotations of molecules are thus generated
as a byproduct of the switching of event-chain directions. In practical
applications, it remains to be seen whether the rotations of molecular
ensembles decay particularly slowly.  In this case only, the ECMC algorithm
will need to be modified in order to explicitly take into account rotations.

\subsection{ECMC for liquid water}
\label{sec:WaterMultipleMolecule}

The SPC/Fw potential is adapted for liquid water at standard temperature
$\SI{300}{K}$ and density $\SI{1}{g/cm^3}$. An ECMC simulation at these
conditions is easily set up with factors (including the dipole--dipole Coulomb
factor) as in \eq{equ:FactorizationSchemeWater} generalized for $N_{\hoh}
>2$. The \quot{ratio}, \quot{outside-first}, and \quot{inside-first}
lifting schemes are taken over from the dipole case discussed in
\sect{sec:Dipoles}. However, the dipole is now constructed from three
particles. For a far distant pair of \hoh\ molecules, the factor
derivatives with respect to the hydrogen positions are usually of the same 
sign, and of opposite
sign to that of the oxygen.  In the notations of \fig{fig:MolFactors} and using
$M=\FACTOR{\SET{1\TO 6}}{\TYPE{Coulomb}}$, we thus have that to order $1/r^3$:
\begin{equation}
\partpartshort{U_M}{x_1} \sim \partpartshort{U_M}{x_3} \sim - \half 
\partpartshort{U_M}{x_2}. 
\end{equation}
This can again be used in the inside-first lifting scheme to keep most of the
lifting flow inside the molecule of the active particle. Care must be exercised
in these lifting schemes to arrange the particles in a fixed order that is
independent of which particle is active (it is incorrect to place the active
particle systematically on the left-most position on the upper row of the table
in \fig{fig:FactorsDipoles}).

\begin{figure}[htb]
\includegraphics[width=\columnwidth]{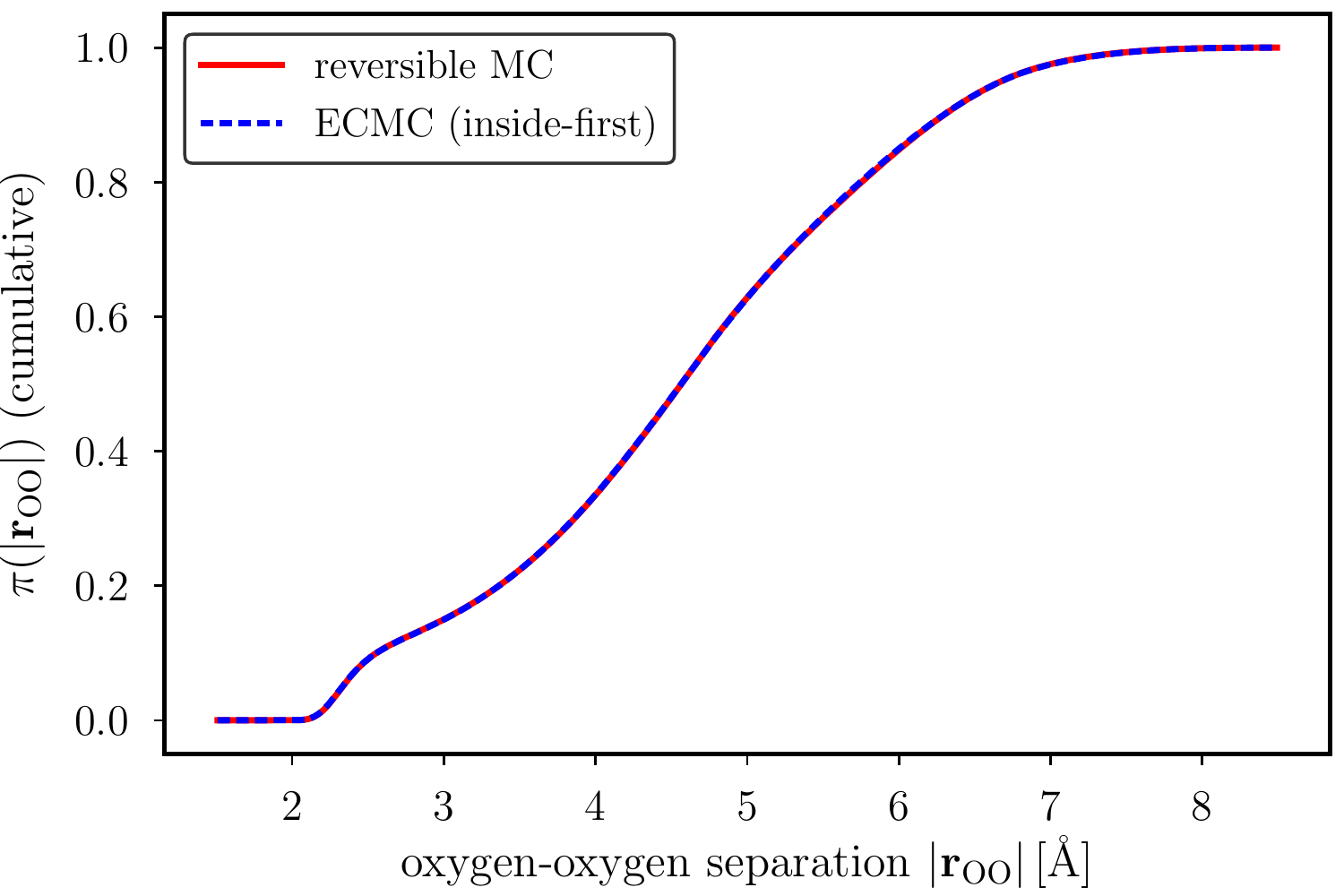}
\caption{Cumulative histogram of the oxygen--oxygen separation $|
\rvec_{\text{OO}}|$ for $32$ \hoh\ molecules at standard density and
temperature via conventional reversible Monte Carlo and ECMC using the factor 
set of \eq{equ:FactorizationSchemeWater} with inside-first lifting scheme.
The random choice of directions was used with a fixed value of  $\ell =
0.5\text{\AA}$. }
\label{fig:OxygenSeparation}
\end{figure}

For long simulation times, the ECMC algorithm exactly samples the Boltzmann
distribution of this model, and thermodynamic observables can be compared with
Metropolis Monte Carlo using the Ewald summation for the Coulomb potential.
This can be verified for the oxygen--oxygen distances that agree to very
high precision, demonstrating that the irreversible ECMC converges towards the 
same steady state as reversible Monte Carlo algorithms
(see \fig{fig:OxygenSeparation}). To make sure that equilibrium is reached, the
initial configurations where chosen randomly in a very dilute system and slowly
compressed towards the target density.

\begin{figure}[htb]
\includegraphics[width=\columnwidth]{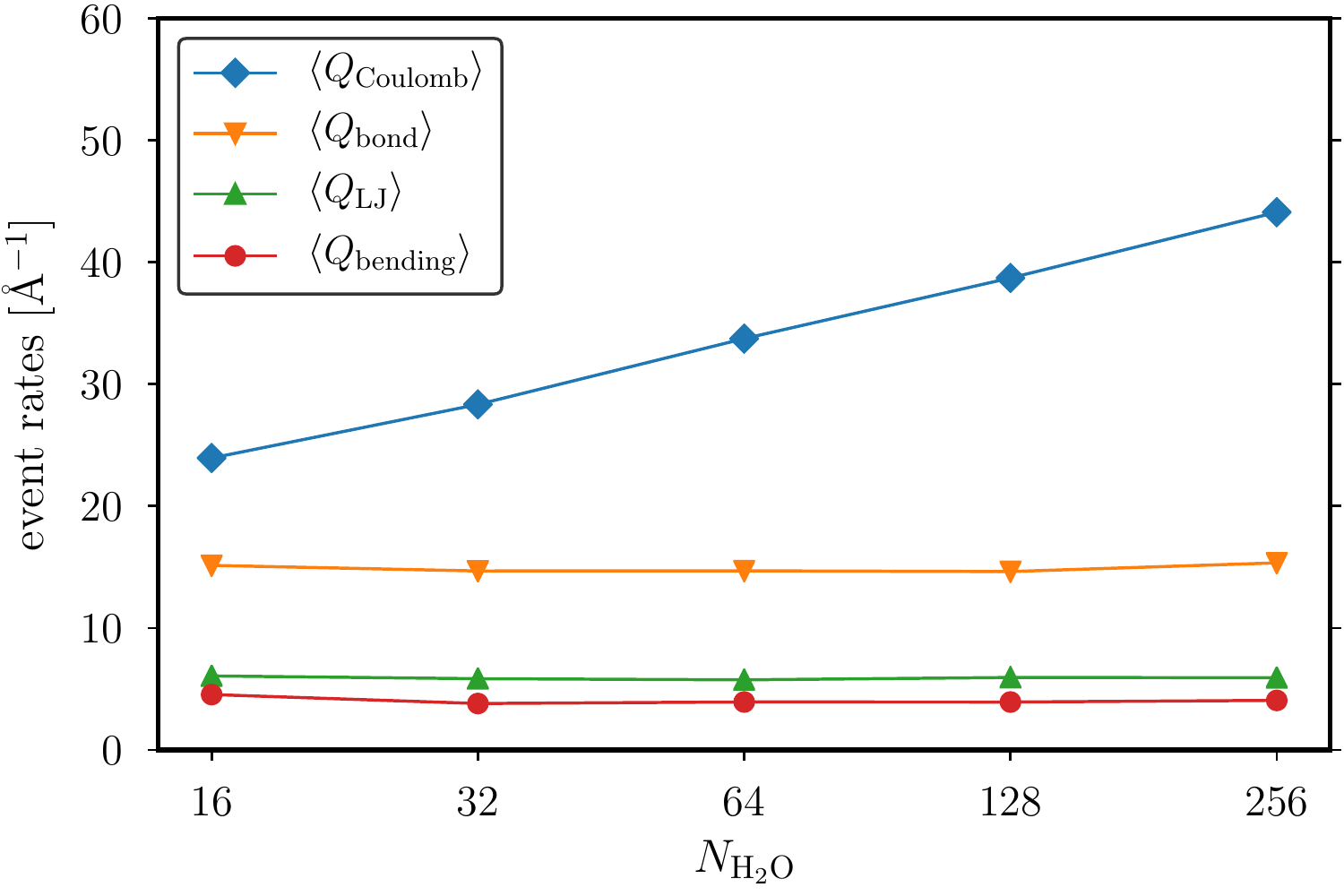}
\caption{Ensemble-averaged total \quot{Coulomb}, \quot{bond}, \quot{LJ}, and 
\quot{bending} event rates as a function of the number of \hoh\ molecules. The 
Coulomb event rate scales logarithmically. Event rates depend on 
the choice of factors but are
independent of the lifting scheme.}
\label{fig:WaterEventRate}
\end{figure}

In the liquid-water simulation for $N_{\hoh} > 2$, the factors $M =
\FACTOR{I_M}{T_M}$ belong to four different types (that is, $T_M \in \TCAL$
and $|\TCAL| =4$), into which the ensemble-averaged total event rate 
with respect to particle $k$ (see \eq{equ:TotalEventRate}) can be split:
\begin{multline}
\mean{Q_k(\SET{\rvec_1 \TO \rvec_N})}  \\ =
\sum_{M \in \SETOFM}
\mean{\EVRATE{M}{k}(\SET{\rvec_i: i \in I_M})} = 
\sum_{T \in \TCAL} \mean {Q_T }.
\label{equ:TotalEventRateSplit}
\end{multline}
$\mean{Q_{\text{Coulomb}}}$ agrees with the definition in \sect{sec:Coulomb}
(see \eq{equ:LogN}). The three local factor types naturally give constant
scaling  of their associated mean event rates $\mean{Q_{\text{bond}}}$,
$\mean{Q_{\text{LJ}}}$, and $\mean{Q_{\text{bending}}}$ with system size,
whereas $\mean{Q_{\text{Coulomb}}}$ clearly features $\log N_{\hoh}$ scaling
with the number of \hoh\ molecules (see \fig{fig:WaterEventRate}). The
logarithmic scaling of the total Coulomb event rate validates the  prediction
of \eq{equ:LogN2}. The total event rate increases by $\SI{5}{\angstrom}^{-1}$
when $N_\hoh$ doubles.

\begin{figure*}[t]
\includegraphics[width=0.8\linewidth]{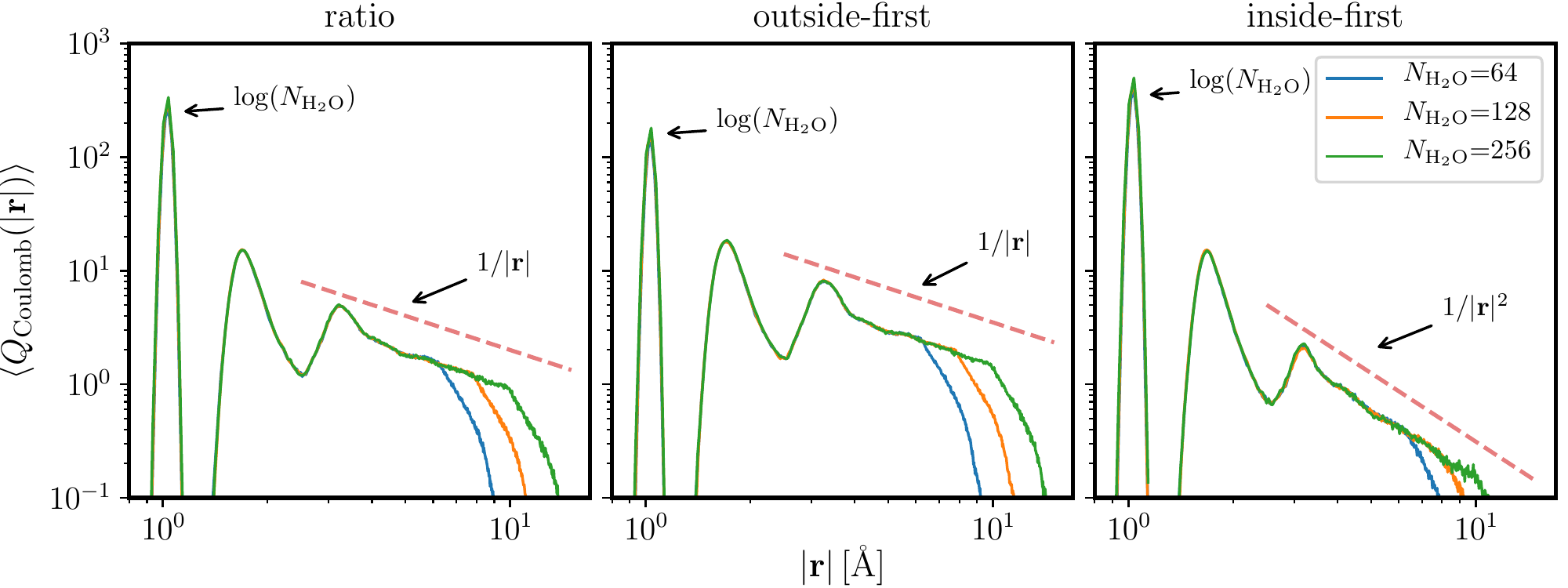}
\caption{Histogram of distance $|\rvec|$ between the active and the target 
particle for
the \COULOMBTYPE\ events for the  \quot{ratio}, \quot{outside-first} and
\quot{inside-first} lifting schemes and for 64, 128 and 256 \hoh\ molecules.
The integral of each histogram corresponds to $\mean{Q_{\text{Coulomb}}}$
in \fig{fig:WaterEventRate}. Dashed lines indicate $| \rvec|^{-z}$ with 
exponents $z$: $1, 1, 2$, which corresponds to  $ q_{\text{inter}} 
\sim |\rvec|^{-(z+2)}$ in Table~\ref{tab:TabFourBody}.}
\label{fig:WaterEventDistance}
\end{figure*}

Finally we study the lifting flows for the Couloequ:FactorPotenialWaterBondmb 
dipole--dipole factors under
the \quot{ratio}, \quot{inside-first}, and \quot{outside-first} schemes (see
\sect{sec:DipoleLiftingSchemes}). As discussed in \sect{sec:DipoleSystemModel},
the event rates are independent of the lifting schemes for a given
factor set. However, the probability distributions of the distance $|\rvec|$
between the active and the target particles are different (see
\fig{fig:WaterEventDistance}).  First, the peak at the oxygen-hydrogen bond
length corresponding to a lifting within the molecule increases logarithmically
with system size.  Second, with increasing system size
the distribution of event distances develops a power-law tail.  In both the
\quot{ratio} and the \quot{outside-first} lifting schemes, the tail of the
probability distribution decreases as $|\rvec|^{-1}$. The
\quot{inside-first} scheme decays as $|\rvec|^{-2}$. These
results, corresponding to the evolution of $ q_{\text{inter}} $ in 
$|\rvec|^{-3}$ and $|\rvec|^{-4}$ in Table~\ref{tab:TabFourBody}.

Remarkably,  the \quot{inside-first} lifting scheme induces mostly local lifting 
flows, even for Coulomb factors that associate \hoh\ molecules that are far 
distant from one another. Most of the liftings are local, and the central peak 
increases as $\log N_\hoh$. We expect a local lifting to keep the dynamics of 
the system coherent, and to lead to faster convergence towards equilibrium.
It appears also possible to replace the interaction with far-away \hoh\ 
molecules by the interaction with an effective medium (given that the lifting 
flow remains local). In the \quot{ratio} and \quot{outside-first} lifting 
schemes, this would probably not be possible as the lifting flow towards 
far-away dipoles is of the same order of magnitude as the local flow. 

\section{Conclusions}
\label{sec:conclusions}

In this work we have outlined the ECMC framework for all-atom computations.
Our algorithm advances a single particle in the presence of long-ranged
electrostatic interactions in \bigO{1} operations, with a mean free path
which decreases as \bigO{1/\loga{N}}.  This gives an overall complexity of
\bigO{N \loga{N}} to advance $N$ particles, each by \bigO{1}. This speed
can be achieved for locally charge-neutral systems, where particles can
be grouped into local dipoles.  The algorithm can take into account the
presence of free point charges, and its performance worsens only gradually
with their number. The algorithm is manifestly translation-invariant and
event-driven. It is free of discretization errors, and exactly samples the
Boltzmann distribution, without needing a thermostat. Its outstanding property
is that it neither computes total forces nor determines the system potential.

ECMC breaks with tradition in two ways.  Firstly, as a Markov-chain algorithm,
it offers the freedom to choose among a variety of moves.  Our approach
of advancing single particles may be a first step only. Nevertheless, as
we have shown, it effectively rotates dipoles and flexible water molecules
in three-dimensional space and samples the entire configuration space. We
have explored the great freedom to choose factors and liftings that suit the
problem at hand. Secondly, ECMC breaks with tradition in that it is purely
Particle--Particle: It treats electrostatic interactions between point charges,
but is oblivious to the electrostatic field. This aspect liberates it from
the interpolating mesh that in traditional Particle--Particle--Particle--Mesh
methods approximates the Coulomb field. Rather, the algorithm is based on the
interaction of pairs of particles and, more generally, of factors that may
comprise pairs of local dipoles or even more complex objects.

In this work, we have checked that thermodynamic quantities from ECMC agree
with those obtained with methods that satisfy detailed balance.  As a next
step for analyzing ECMC in all-atom systems, it will be important to study
its relaxation dynamics in detail. This dynamics will certainly depend on the
choice of factors and, for example, for the case of dipole--dipole factors
treated here, on the choice of liftings. The inside-first lifting scheme
yields mostly local dynamics, and we would expect it to lead to a faster
decay of correlation functions. Besides this, we have discussed that the
length and the probability distribution of the event-chain parameter $\ell$,
and even the sequence of the directions of the event-chain can significantly
influence the ECMC dynamics although, as we have verified extensively, the
steady state is always given by the Boltzmann distribution.  We would hope
that, in addition to the overall favorable algorithmic scaling, the fast decay
of density fluctuations carries over from short-range-interacting particle and
spin systems.  The influence of different factorization and lifting schemes on
the dynamics of ECMC will also have to be understood.  From an algorithmic
implementation point of view, we think that the parallelization of the
method~\cite{KapferPolytope2013} will have to be dealt with carefully.

\begin{acknowledgments}
M.F.F. acknowledges financial support from EPSRC fellowship EP/P033830/1 and
hospitality at Ecole normale sup\'{e}rieure.  This work was initiated at the
Aspen Center for Physics, which is supported by National Science Foundation
grant PHY-1066293. We thank Matthew Downton for illuminating discussions and
Matthias Staudacher for helpful comments.  W.K. acknowledges support from
the Alexander von Humboldt Foundation and thanks the Santa Fe Institute for
hospitality.
\end{acknowledgments}

\bibliography{../Factsheet/General,../Factsheet/Water}
\end{document}